\documentclass{article}
\usepackage{graphicx} 
\usepackage[a4paper, margin=2.5cm]{geometry}
\usepackage{algpseudocode}
\usepackage{amsmath}
\usepackage{amssymb}
\usepackage[title]{appendix}
\usepackage[labelfont=bf]{caption}
\usepackage{enumitem}
\usepackage[colorlinks, citecolor=blue]{hyperref}
\usepackage{multirow}
\usepackage{natbib}
\usepackage{lscape}
\usepackage{placeins}
\usepackage{subcaption}
\usepackage{tabularx}
\usepackage{titlesec}
\usepackage{tabularx}
\usepackage{array}
\usepackage{threeparttable}

\usepackage{array}
\newcolumntype{C}[1]{>{\centering\arraybackslash}p{#1}}

\setcitestyle{citesep={,}}
\titlespacing*{\section}{0pt}{30pt}{20pt}

\algdef{SE}[LOOP]{Loop}{EndLoop}[1]{\textbf{Loop} #1 \textbf{Do}}{\textbf{End Loop}}
\algdef{SE}[PROCEDURE]{Procedure}{EndProcedure}[2]{\textbf{Procedure} #1(#2)}{\textbf{End Procedure}}

\title{Grow and Pollute but Invest and Clean: Dynamic Associations between Parent Firm Characteristics and Facility Toxic Releases}
\author{George Kapetanios \thanks{King’s Business School, King’s College London, 30 Aldwych, London, WC2B 4BG (United Kingdom); Email: george.kapetanios@kcl.ac.uk.} \and Steven Ongena \thanks{Department of Finance, University of Zurich, Plattenstr. 14, 8032 Zurich (Switzerland); Swiss Finance Institute, KU Leuven, NTNU Business School and CEPR; Email: steven.ongena@df.uzh.ch.} \and Alexia Ventouri \thanks{(Corresponding author): King’s Business School, King’s College London, 30 Aldwych, London, WC2B 4BG (United Kingdom); Tel: +44-2078-483132, Email: alexia.ventouri@kcl.ac.uk.} \and Huiyan Xiao \thanks{King’s Business School, King’s College London, 30 Aldwych, London, WC2B 4BG (United Kingdom); Email: huiyan.xiao@kcl.ac.uk} }

\begin{document}

\begin{titlepage}

\date{}
\maketitle

\vspace*{3cm}

\begin{abstract}

This paper examines how relationships between parent-firm characteristics and facility-level toxic releases evolve over time. Using 238,304 observations for 7,447 U.S. manufacturing facilities from 1992 to 2023, we link on-site releases from the Toxics Release Inventory to financial, managerial, and macroeconomic data. A time-varying mean-group estimator accommodates changes in average coefficients over time and heterogeneity across facilities. The main result is a persistent contrast between operating scale and investment-related adjustment: sales is positively associated with release growth, whereas investment intensity is negatively associated. This pattern remains in joint specifications, parsimonious models, and the main robustness exercises. Other financial and managerial associations are identified as well. These findings suggest that environmental policy evaluation may benefit from distinguishing release changes associated with production expansion from those associated with investment-related adjustment and from allowing for variation across periods and production settings. For corporate managers, the same distinction can inform how emissions considerations are incorporated into capital budgeting and production planning.

\end{abstract}

\vspace{1em}

\end{titlepage}

\newpage

\section{Introduction}

Over the past three decades, environmental regulation and public scrutiny have transformed the way firms internalize pollution externalities \citep{becker2000effects,greenstone2002impacts,greenstone2003estimating,shapiro2018pollution,gaganis2021informal}. What was once treated primarily as a compliance cost is increasingly embedded in firms' investment strategies, organizational capabilities, and innovation decisions. Emissions outcomes today reflect not only regulatory stringency, but also how firms adapt technologically and financially to evolving policy regimes and macroeconomic conditions. Understanding these adaptive processes is central to debates in innovation and industrial policy, where regulation is often viewed not merely as a constraint, but as a potential catalyst for technological change and organizational learning \citep{porter1995toward}.

A growing body of research argues that environmental policy can stimulate innovation by altering relative prices, inducing process improvements, and accelerating capital-vintage turnover \citep{porter1995toward,shapiro2018pollution}. Meanwhile, firms differ markedly in their ability to respond. Financial structure, investment capacity, and managerial characteristics shape whether regulation leads to technological upgrading, organizational change, or delayed compliance \citep{konar2001does,tomar2023greenhouse}. Recent evidence further highlights the role of intrinsic motivation and place-based preferences in shaping corporate pollution abatement, showing that adaptation to environmental challenges is mediated by internal organizational factors as well as external policy constraints \citep{andrikogiannopoulou2025not}.

These heterogeneous responses imply that the relationship between firm characteristics and emissions is unlikely to be stable over time. Instead, it should evolve as regulatory regimes mature, technologies diffuse, and macroeconomic conditions change. Despite this, much of the empirical literature still relies on static models that impose constant effects over long horizons or focus on discrete policy shocks. Such approaches obscure the dynamic nature of firm adaptation. A factor that is strongly associated with emissions in an early regulatory phase, such as scale or investment intensity, may weaken, reverse, or disappear as firms learn, innovate, or face new financial constraints. Capturing these evolving relationships is therefore important for understanding how production scale, investment, and organizational characteristics are associated with emissions adjustment over time.

In order to better understand these dynamic developments, this paper studies the dynamic evolution of emissions correlates using a long and granular panel of U.S. manufacturing facilities, drawn from the Toxics Release Inventory (TRI). The broad balanced facility file contains a quarter million facility-year observations, covering three decades, more than 7,000 facilities operated by more than 300 publicly-listed firms, with annual emissions from 1992 onwards. We link facility-level on-site releases to parent-firm financial characteristics from Compustat, managerial attributes from ExecuComp, and aggregate macroeconomic indicators from FRED-QD. The matched data primarily describe facilities owned by listed parent firms in the selected manufacturing industries.

We employ a time-varying mean-group (TVMG) estimator that allows regression coefficients to evolve smoothly over time while accommodating heterogeneity across facilities \citep{robinson1989nonparametric,robinson1991time,pesaran1995estimating,giraitis2014inference,giraitis2018inference,giraitis2021time}. The approach combines kernel-weighted local estimation at the unit level with cross-sectional aggregation, yielding a time path of mean coefficients. This framework is well suited to documenting gradual changes in the associations between firm characteristics and emissions without prespecifying breakpoints or policy dates.

Our results reveal notable time variation, but they also show a persistent contrast between operating scale and investment-related adjustment. Parent-firm sales is positively associated with facility emissions growth, whereas investment intensity is negatively associated with emissions growth for most periods. These contrasting findings remain present in both augmented joint and more parsimonious models. They also remain when releases are aggregated to the parent-firm level. Other financial, valuation, and managerial associations are identified as well, with significance and directions less stable across samples and specifications. The additional exercises support and refine this central result. The sales and investment pattern remains evident with simultaneous confidence bands, after removing inserted facility-years with no TRI record, and under alternative smoothing choices, as well as other robustness settings. Sectoral estimates show that the timing and direction of the investment association vary across production settings, while positive sales associations are more widespread. For example, investment intensity is negatively associated with emissions growth in machinery but positively associated with it in printing, consistent with sectoral differences in whether capital spending accompanies process adjustment or capacity expansion. The foreign operation exposure does not account for the main pattern of associations in domestic releases. Aggregate economic factors co-move with changes in total releases and some specific channels.

We also examine whether the negative investment association supports a causal interpretation. The proposed interaction between debt-maturity exposure and the Excess Bond Premium is weak in pooled and local specifications, and the alternative instrument designs do not materially improve identification. Weak-instrument-robust Anderson--Rubin confidence sets do not identify a finite causal direction or magnitude. The Monte Carlo exercises assess the coverage of the inference procedure under simulated designs, while the strength and validity of the empirical instrument remain separate empirical questions. We therefore carefully interpret the TVMG estimates as time-varying conditional associations rather than causal effects of investment or environmental regulation.

These findings contribute in three ways to the academic and policy literature. First, they provide long-run evidence on how relationships between firm characteristics and industrial emissions evolve across different economic environments, reinforcing the link between innovation policy, access to finance, and organizational decision-making. Second, they document a persistent distinction between operating scale and investment-related adjustment in their association with toxic release and show that these associations survive joint estimation, changes in the unit of analysis, reported production/activity controls, and the main robustness exercises. Third, the paper provides a TVMG framework for studying evolving relationships between emissions and firm characteristics in heterogeneous panel data. By allowing coefficients to vary over time and across units, the framework can be applied to related environmental and corporate panel settings.

From a policy and corporate-management perspective, these patterns suggest several considerations for innovation-oriented environmental regulation and firm decision-making. The variation in the estimated associations across periods indicates that policy evaluation may need to allow for changes in firms' financial positions and the macroeconomic environment rather than presume uniform relationships over time. For corporate managers, the same distinction offers a practical lens for integrating emissions information into capital budgeting and production planning. Firms may find it useful to distinguish capital expenditure aimed primarily at capacity expansion from expenditure on replacement or process improvement, and to reassess that composition as financing conditions and production environments change.

To provide a disciplined interpretation of these time-varying reduced-form estimates, we also draw on a simple economic framework in which total emissions depend on both production scale and emissions intensity, while firms can lower future emissions intensity through environmentally oriented investment whose effective cost depends on a broad set of firm-level and external conditions. This interpretation is consistent with evidence that environmental performance affects firm value and financing conditions, and with a growing climate-finance literature showing that investor demand and the pricing of climate risk have become increasingly salient \citep{konar2001does,chava2014environmental,krueger2020importance,bolton2021do,pastor2021sustainable}. The purpose of this framework is not to identify structural parameters. Rather, it provides a coherent way to understand why the association between emissions growth and variables such as sales, investment intensity, cash holdings, and macroeconomic conditions may itself evolve over time.

The remainder of the paper proceeds as follows. Section~\ref{sec:model} presents the economic framework, TVMG estimator, and data. Section~\ref{sec:result} reports the baseline results, robustness exercises, and joint model estimations. Section~\ref{sec:conclusion} concludes.

\section{Time-varying Mean-group Model for Emission Analysis}    \label{sec:model}

\subsection{Economic framework}

To interpret empirical estimates, it is important to have an economic framework that guides the interpretation of results.  
It is useful to start by distinguishing between a scale margin and an adjustment margin. Suppressing firm subscripts for expositional clarity, let total emissions be denoted by $\mathcal{E}_t$ and write
\begin{equation}
    \mathcal{E}_t = m_t q_t,
    \label{equ:cf_emissions}
\end{equation}
where $q_t$ denotes current activity and $m_t$ denotes emissions intensity, that is, emissions per unit of activity. Taking log differences gives
\begin{equation}
    \Delta \log \mathcal{E}_t = \Delta \log m_t + \Delta \log q_t.
    \label{equ:cf_growth}
\end{equation}
Equation \eqref{equ:cf_growth} provides the basic economic lens for the empirical analysis. Emissions growth may be high because activity expands, because emissions intensity rises, or because emissions intensity does not decline sufficiently quickly.

We assume that firms can lower future emissions intensity through environmentally oriented investment, which may include cleaner capital replacement, pollution-control equipment, process redesign, or other forms of operational adjustment. Let $G_t$ denote such investment. Emissions intensity then evolves according to
\begin{equation}
    m_{t+1} = (1-\rho)m_t - \chi G_t + \varepsilon_{t+1},
    \qquad 0 < \rho < 1,\quad \chi > 0,
    \label{equ:cf_intensity}
\end{equation}
so that higher environmentally oriented investment reduces next period's emissions intensity. This interpretation is consistent with evidence that long-run declines in industrial pollution are driven to an important extent by falling emissions intensity rather than by output contraction alone \citep{shapiro2018pollution}.

A simple way to represent the cost of environmental adjustment is to assume that firms choose current activity $q_t$ and environmentally oriented investment $G_t$ to maximize
\begin{equation}
    R(q_t;A_t) - C(q_t) - \lambda_t m_t q_t - \frac{1}{2}(\kappa + \psi_t)G_t^2 + \delta \mathbb{E}_t[V(m_{t+1})],
    \label{equ:cf_problem}
\end{equation}
subject to \eqref{equ:cf_intensity}. Here $A_t$ is an aggregate demand or productivity shifter, $\lambda_t$ is the effective shadow cost of emissions, and $\psi_t$ is a composite wedge that captures factors affecting the cost of environmentally oriented investment. We write
\begin{equation}
    \psi_t = \Psi(\mathcal{F}_t),
    \qquad \kappa + \psi_t > 0,
    \label{equ:cf_wedge}
\end{equation}
where $\mathcal{F}_t$ denotes the set of firm-level and external conditions that may affect investment costs. It can include balance-sheet characteristics such as assets and leverage, liquidity, aggregate financial conditions, and other relevant conditions. We do not impose a particular functional form or assign a structural sign to any individual empirical proxy.

This setup also clarifies the interpretation of the observed investment-intensity variable used in the empirical analysis. Let total observed investment be denoted by $I_t$, and let the share of that investment directed toward emissions-reducing adjustment be $\omega_t \in [0,1]$. Then
\begin{equation}
    G_t = \omega_t I_t = \omega_t K_{t-1} \text{Invint}_t,
    \label{equ:cf_green_share}
\end{equation}
where $\text{Invint}_t = I_t/K_{t-1}$ is observed investment intensity and $K_{t-1}$ is a scale variable such as lagged PP\&E. Equation \eqref{equ:cf_green_share} implies that the reduced-form coefficient on investment intensity need not be constant over time. When the environmentally oriented share of investment is small, investment intensity mainly reflects ordinary expansionary investment; when that share rises, the same observed variable increasingly proxies emissions-reducing adjustment. For the investment variable in our study, we do not claim a fix part as ``green" investments, but broad capital expenditure, as the true amount put to environmental compliance is hard to observe.

The framework yields four qualitative implications that are directly relevant for the empirical analysis. First, holding emissions intensity fixed, higher activity raises emissions, so variables that proxy current scale, such as sales, are more likely to be positively associated with emissions growth when the scale channel dominates. Second, greater environmentally oriented investment lowers future emissions intensity and therefore tends to reduce subsequent emissions growth. Third, firm-level and external conditions can alter the cost of such adjustment through the composite wedge $\psi_t$, without implying a fixed structural sign for assets, leverage, cash holdings, or any other individual proxy. Fourth, because the state variables $(A_t,\lambda_t,\mathcal{F}_t,\omega_t)$ evolve over time, the reduced-form relationship between emissions growth and firm characteristics need not be stable across periods. The role of the empirical specification is therefore not to estimate structural parameters of \eqref{equ:cf_problem}, but to document how the visibility of these economic margins changes over time. The appendix provides more detail on this framework.

\subsection{Methodology}

Following \citet{giraitis2014inference} and \citet{giraitis2021time}, we specify the nonparametric time-varying model as
\begin{equation}
    y_{it} = x'_{it} \beta_{it} + u_{it}, \quad i = 1, ..., N, \quad t = 1, ..., T,
\end{equation}
where $y_{it}$ is the outcome, $x_{it}=(x_{1,it}, ..., x_{p,it})'$ is a $p \times 1$ vector of explanatory variables, $\beta_{it}=(\beta_{1,it}, ..., \beta_{p,it})'$ is a $p \times 1$ coefficient vector that may vary over time, and $u_{it}$ is a scalar error term.

Following \citet{giraitis2021time} and \citet{bai2023mean}, the unit-specific coefficient at evaluation date $t$ is estimated as
\begin{equation}
    \hat{\beta}_{it} = \left(\sum_{j=1}^{T} b_{H,|j-t|} x_{ij} x_{ij}' \right)^{-1}\left( \sum_{j=1}^{T}b_{H,|j-t|} x_{ij} y_{ij} \right),
\end{equation}
where the kernel weights are
\begin{equation}
    b_{H,|j-t|} = K \left( \frac{|j-t|}{H} \right),
\end{equation}
and $H$ controls the degree of temporal smoothing. The kernel $K(\cdot)$ maps the scaled distance from the evaluation date into a nonnegative weight. We assume that $K(\cdot)$ is continuous, has either compact support or sufficiently fast tail decay, and that $K$ and $K'$ are bounded by $C(1+|x|^v)^{-1}$ for some $C>0$ and $v\geq2$. Common examples include the Epanechnikov and Gaussian kernels. The conditions $H\rightarrow\infty$ and $H=o(T)$ as $T\rightarrow\infty$ balance precision against temporal localization: the effective sample around $t$ grows while the estimation window remains local. For a Gaussian kernel, $K(x)\propto\exp(-cx^\alpha)$.

We use a mean-group estimator rather than pooled OLS. At each date $t$, the mean coefficient is the simple cross-sectional average of the unit-specific estimates:
\begin{equation}
    \hat{\beta}_{MG,t} = \frac{1}{N}\sum^{N}_{i=1}\hat{\beta}_{it},
    \label{equ:mg}
\end{equation}
This construction follows the mean-group logic of \citet{pesaran1995estimating} and parallels \citet{bai2023mean} without the instrumental-variable component. Each unit-specific regression includes an intercept.

\paragraph{Remark 1.}
All unit-specific local regressions include a constant, $x_{1,it}=1$. Its time-varying coefficient, $\beta_{1,it}$, serves as a unit effect $\alpha_{it}$ that may drift slowly over $t$ under the same smoothness conditions as the slope coefficients. An equivalent approach would first remove an intercept within each unit, using either a time-invariant intercept when $\alpha_i$ is constant or a kernel-smoothed intercept when $\alpha_{it}$ varies, and then estimate the TVMG regression on the residuals. Under the maintained assumptions, this preliminary step does not change the theoretical analysis or inference for the slope paths. Including the intercept directly accommodates slowly evolving facility baselines without an additional filtering step; we do not interpret the intercept path itself.
\paragraph{}

Next, we outline relevant assumptions on $x_{it}$ and $u_{it}$, following \citet{giraitis2014inference}, \citet{giraitis2021time}, and \citet{bai2023mean}.

\paragraph{Assumption 1.}
Elements of $x_{it}$ and $u_{it}$ have the following properties.
\begin{enumerate}[label=(\roman*)]
    \item There exist $\theta > 4$ and a constant $C < \infty$ such that, uniformly over the relevant indices,
    \begin{equation*}
        E|x_{\ell,it}|^\theta \le C, \quad E|u_{it}|^\theta \le C.
    \end{equation*}
    
    \item For each unit $i$ and regressor component $\ell$, the mean-centered processes $\{x_{\ell,it}-Ex_{\ell,it}\}$ and $\{u_{it}\}$ are strongly mixing ($\alpha$-mixing), with coefficients $\alpha^{i,j}_k$ satisfying, for $0<\phi_{i,j}<1$, $c_{i,j}>0$, and $k\geq1$,
    \begin{equation*}
        \alpha^{i,j}_k \le c_{i,j}\phi^k_{i,j}, \quad  j \in \{x, u\}.
    \end{equation*}

    \item $E[u_{it} | x_{it}] = 0$.

    \item For each unit $i$, let $\Sigma^i_{xx,t} = E[x_{it}x'_{it}]$. Then
    \begin{equation*}
        \max_{t \ge 1} \parallel (\Sigma^i_{xx,t})^{-1} \parallel_{sp} < \infty.
    \end{equation*}
\end{enumerate}

\paragraph{Assumption 2.}
Let $e_i$ denote the random-coefficient deviation in $\beta_i=\beta_0+e_i$. The vectors $(x_i,u_i,e_i)$ are mutually independent across $i$, where $x_i=(x'_{i1},x'_{i2},...,x'_{iT})'$, $u_i=(u_{i1},u_{i2},...,u_{iT})'$, and $e_i=(e_{i1},e_{i2},...,e_{iT})'$.

\paragraph{Assumption 3.}
The coefficients $\beta_{it}$ follow the random-coefficient model
\begin{equation*}
    \beta_{it} = \beta_{0,t} + e_{it}, \quad i = 1, ..., N, \quad t = 1, ..., T,
\end{equation*}
with the following properties:
\begin{enumerate}[label=(\roman*)]
    \item $\beta_{0,t}=E(\beta_{it})$ is the nonrandom sequence of time-varying cross-sectional mean coefficients.

    \item For each component $\ell$, $\beta_{0,\ell,t}$ is uniformly bounded in $t$ and satisfies the smoothness condition
    \begin{equation*}
    |\beta_{0,\ell,t} - \beta_{0,\ell,s}| \le C(\frac{|t - s|}{T})^{\gamma_1}, \quad t,s = 1, ..., T,
    \end{equation*}
    for some $0 < \gamma_1 \le 1$ and a positive constant $C < \infty$ independent of $\ell$, $t$, $s$, $T$.

    \item For each $i$ and $\ell$, the random-coefficient component $e_{\ell,it}$ satisfies the smoothness condition
    \begin{equation*}
        |e_{\ell,it} - e_{\ell,is}| \le (\frac{|t - s|}{T})^{\gamma_2}q_{\ell,i,ts}, \quad t,s = 1, ..., T,
    \end{equation*}
    for some $0<\gamma_2\leq1$. Each random variable $X\in\{e_{\ell,it},q_{\ell,i,ts}\}$ satisfies the thin-tail condition $\varepsilon(\alpha)$: for all $\omega>0$,
    \begin{equation*}
        \Pr(|X| \ge \omega) \le \exp(-c_1|\omega|^\alpha),
    \end{equation*}
    where $c_1>0$ and $\alpha>0$ are independent of $\ell$, $t$, $s$, and $T$.
\end{enumerate}
\paragraph{}

Assumption 1(i) imposes high-order moment bounds on the regressors and errors. These bounds control the tails of the kernel-weighted sums used in the unit-level local regressions. Uniformly bounded moments of order $\theta>4$ permit exponential and maximal inequalities that support uniform consistency and pointwise asymptotic normality at interior dates. This condition is compatible with firm-level variables after standard scaling or logarithmic transformation. Assumption 1(ii) imposes strong-mixing ($\alpha$-mixing) conditions on each time series. It permits serial correlation and conditional heteroskedasticity but rules out long memory. Geometric decay of the mixing coefficients supports laws of large numbers and central limit theorems for the kernel averages. Assumption 1(iii) supplies the exogeneity condition for local OLS: at each date $t$, the regressors are mean-independent of the disturbance, so the local estimator targets $\beta_{it}$. Assumption 1(iv) requires the regressor covariance matrix $\Sigma^i_{xx,t}=E[x_{it}x'_{it}]$ to remain uniformly nonsingular. This condition rules out near-perfect local collinearity and prevents the sampling variance from diverging.

Assumption 2 formalizes the cross-sectional structure of the random-coefficient mean-group model. Each unit's coefficient path is $\beta_i=\beta_0+e_i$, where $\beta_0$ is the deterministic target and $e_i$ captures unit-specific deviations. Independence of $(x_i,u_i,e_i)$ across units permits unrestricted heterogeneity in their marginal distributions and yields a cross-sectional central limit theorem for $\hat{\beta}_{MG}=N^{-1}\sum_i\hat{\beta}_i$. Its asymptotic variance is determined by the cross-sectional dispersion of $e_i$.

Assumption 3 specifies how the coefficients vary over time. Part (i) defines the estimand $\beta_{0,t}=E(\beta_{it})$, the nonrandom cross-sectional mean coefficient path targeted by the mean-group estimator. Part (ii) imposes H\"older continuity on each component of $\beta_{0,t}$. It bounds the rate of drift and implies deterministic kernel bias of order $O((H/T)^{\gamma_1})$ within a local window of width $H$. Part (iii) allows the unit-specific deviations $e_{it}$ to evolve idiosyncratically but smoothly, subject to an analogous H\"older bound and thin-tailed random moduli $q_{\ell,i,ts}$. Together, these conditions support uniform consistency and pointwise asymptotic normality at interior dates.

\paragraph{Remark 2.}
Under Assumptions 1--3, let $H\rightarrow\infty$ and $H=o(T)$. As $(N,T)\rightarrow\infty$, the estimator is uniformly consistent:
\begin{equation*}
    \max_{t=1, ..., T} \parallel \hat{\beta}_{MG,t} - \beta_{0,t} \parallel = O_p\left( \left(\frac{H}{T}\right)^{\gamma_1} + \sqrt{\frac{\log{T}}{H}} + \frac{1}{\sqrt{N}} \right).
\end{equation*}
For pointwise asymptotic normality, if the bias is negligible relative to $N^{-1/2}$, i.e. $(H/T)^{\gamma_1} = o(N^{-1/2})$, then for any interior time $t$,
\begin{equation*}
    \sqrt{N} (\hat{\beta}_{MG,t} - \beta_{0,t}) \xrightarrow{d} \mathcal{N}(0, \Sigma_{e,t}), \quad \Sigma_{e,t} = \lim_{N\to\infty}\operatorname{Var}\!\left(\frac{1}{\sqrt{N}}\sum_{i=1}^N e_{it}\right).
\end{equation*}
With $\hat\beta_{it}$ denoting the unit-level local OLS estimator at date $t$,
\begin{equation*}
    \hat{\Sigma}_{e,t} = \frac{1}{N}\sum^N_{i=1}(\hat{\beta}_{it} - \hat{\beta}_{MG,t})(\hat{\beta}_{it} - \hat{\beta}_{MG,t})' \xrightarrow{p} \Sigma_{e,t}.
\end{equation*}

\paragraph{Bandwidth selection}

The asymptotic requirements $H\rightarrow\infty$, $H=o(T)$, and $(H/T)^{\gamma_1}=o(N^{-1/2})$ motivate the parameterization $H=T^\alpha$ with $0<\alpha<1$ \citep{giraitis2014inference,giraitis2018inference,giraitis2021time}. Two common approaches select $\alpha$. A leave-one-unit-out cross-validation procedure can choose it from a discrete grid, such as $\alpha\in\{0.3,0.35,...,0.8,0.85\}$ in \citet{bai2023mean}. Alternatively, the rule of thumb $\alpha=1/2$ applies the same degree of smoothing; \citet{giraitis2018inference} and \citet{giraitis2021time} report favorable finite-sample performance for this choice. Cross-validation adapts to each sample but can produce different smoothing levels across variables and can respond to small differences in sample composition. A common square-root bandwidth makes the coefficient paths more directly comparable. We therefore use $H=T^{1/2}$ in most analysis and examine alternative bandwidths and kernels in Section \ref{sec:sensitivity}.
\paragraph{}

Applied to the facility-release data, the model can be written as
\begin{equation}
    y_{it} = \alpha_{it} + x'_{J(i,t)}\beta_{t} + u_{it},
    \label{equ:reg}
\end{equation}
where $J(i,t)$ maps facility $i$ in year $t$ to its parent firm $j$, and $x_{J(i,t)}$ contains the corresponding parent-firm covariates. The parameter $\beta_t$ is the mean-group slope, $\alpha_{it}$ is the unit-specific effect described in Remark 1, and $u_{it}$ is the idiosyncratic error. We first estimate the unit-specific slopes by kernel smoothing and then average them across units as in Equation \eqref{equ:mg}. Following \citet{andrikogiannopoulou2025not} and \citet{greenstone2003estimating}, the outcome is the symmetric percentage change in releases:
\begin{equation}
    \%\Delta Y_{i,t} = \frac{Y_{i,t} - Y_{i,t-1}}{(Y_{i,t} + Y_{i,t-1})/2}.
    \label{equ:emissionchange}
\end{equation}
This measure takes values in $[-2,2]$ and treats expansions and contractions symmetrically. The outcome includes on-site releases through air, water, and ground channels.

\subsection{Data}

Our data combine three sources: facility-level releases and facility characteristics from the U.S. Environmental Protection Agency's Toxics Release Inventory (TRI), parent-firm financial and chief-executive information from Compustat and ExecuComp, and aggregate macroeconomic series from FRED-QD. The annual coverage is 1992--2023.

The merged data are organized at the unique TRI facility-year level. Since we aim to adopt Gaussian kernel smoothing, a broad coverage screen retains facilities observed throughout 1992--2023 and requires a nonmissing parent firm in every estimation year from 1993--2023; the 1992 parent identifier may be missing because that year is used to construct the first emissions change. This screen yields 7,447 facilities and 238,304 facility-year observations. It does not impose common completeness for all financial and CEO variables.

\subsubsection{Facility Data}    \label{sec:fac_data}

Facility-level releases come from the TRI, established under the Emergency Planning and Community Right-to-Know Act of 1986.\footnote{TRI webpage: https://www.epa.gov/toxics-release-inventory-tri-program.} We fix the chemical and industry scope over the full sample to exclude changes in measured emissions that arise from changes in reporting coverage. On the chemical dimension, we retain only releases involving chemicals on the EPA's 1991 Core Chemicals list. \footnote{EPA 1991 Core Chemicals: https://enviro.epa.gov/triexplorer/tri\_text.list\_chemical\_core\_91.} TRI periodically updates and releases its list of core chemicals subject to reporting requirements. To construct a consistent facility-level emissions panel over time, we rely on the core chemicals list applicable to the beginning of our sample period. Specifically, we use the 1991 TRI core chemical list to ensure that the set of chemicals included in the analysis remains comparable across facilities and years, avoiding potential inconsistencies arising from later revisions to the reported chemical coverage. On the industry dimension, we retain facilities whose harmonized primary three-digit NAICS code is in 311--316, 321--327, or 331--339. The scope of TRI reporting requirements has expanded over time. Initially, TRI reporting primarily covered facilities in the manufacturing sector classified as the previously described codes. In 1998, the EPA expanded TRI coverage to include additional industry sectors, which introduced potential inconsistencies in sectoral coverage over time. To maintain comparability throughout the sample period, we restrict our analysis to pre-1998 listed sectors, ensuring that observed changes in facility-level releases are not driven by changes in the reporting scope. The chemical and industry filters are applied to the underlying TRI chemical reports before releases are aggregated to the facility-year level.

Because the analysis fixes both chemical and industry coverage and applies a consistent set of mandatory reporting requirements, year-to-year changes in facility appearance are unlikely to arise mechanically from changes in the statutory sample definition. A facility that disappears from the TRI in a particular year, or enters the data for the first time, may have ceased or begun operations, or may have moved below or above the applicable reporting threshold. Although reporting errors or noncompliance cannot be ruled out entirely, systematic nonreporting is less likely given the regulated and mandatory nature of TRI disclosure. Importantly, this study focuses specifically on toxic releases observable within the TRI reporting framework rather than attempting to measure all physical emissions generated by industrial facilities. Releases associated with activities below the applicable TRI reporting thresholds therefore fall outside the measurement scope of the analysis. To the extent that these omitted releases are concentrated among relatively low-activity facilities and are small compared with releases reported above the thresholds, their effect on the aggregate results is expected to be limited. Given this, we assign ``zero emissions" to an otherwise eligible facility that exists in some years but does not appear in a given year. By doing this, we integrate shut-down and new-establishment of facilities, which should be considered as having non-excusable impact on emissions changes. The resulting value represents zero contribution to observed TRI releases in that year, rather than a verified absence of all physical emissions.

The outcome includes on-site releases only. Air releases combine fugitive and stack air; water releases cover discharges to surface water; and ground releases combine underground injection, land disposal, surface impoundments, and other on-site disposal. Total releases equal air plus water plus ground. Recycling, energy recovery, treatment, and off-site transfers are excluded. Amounts are expressed in pounds, with TRI forms reported in grams converted to pounds before aggregation. Two consecutive zero-release years are assigned a change of zero, and the first usable outcome year is 1993.

TRI facility-year observations were matched to Compustat firms using parent-company names within the same adjusted reporting year. To align fiscal and calendar years, the matching year was defined as the datadate year minus one for firms with fiscal year ends between January and May and as the datadate year otherwise. Company names were converted to uppercase ASCII text, stripped of punctuation, leading articles, conjunctions, and trailing legal-form terms such as “Inc.,” “Corp.,” “LLC,” and “Ltd.,” and the remaining words were sorted to allow exact matches despite differences in word order. Each of four TRI parent-name fields—the reported and standardized domestic and foreign parent names—was matched independently against both Compustat’s abbreviated and full company-name fields.

Most facility exclusions arise because their parent companies cannot be matched to Compustat, which primarily covers publicly listed firms. These unmatched facilities are therefore likely to be owned by private or otherwise non-listed companies, although the absence of a GVKEY may also reflect missing or ambiguous parent-name information. Accordingly, the empirical results should be interpreted as describing the emissions behavior of facilities owned by listed firms rather than the entire population of TRI-reporting facilities.

\subsubsection{Firm-Level Financial and CEO Data}

Financial data are obtained from Compustat North America Fundamentals Annual, and CEO data are obtained from ExecuComp Annual Compensation; both sources are provided by S\&P Global Market Intelligence. Compustat supplies the accounting inputs used to measure size, financing, investment, and valuation. ExecuComp supplies CEO identity, age, and appointment information. When a firm-year contains two CEOs, we average their available ages.

The outcome measures release growth from year $t-1$ to year $t$. To align covariate timing, assets, leverage, investment intensity, and cash holdings enter with the lags shown in Table \ref{tab:variable_definition}, while sales, Tobin's Q, and CEO characteristics are measured in year $t$. Lagged financial variables are constructed by matching the same GVKEY in the required earlier calendar year; the previous available observation is not substituted when a year is missing. CEO tenure equals the calendar year minus the year in which the executive became CEO. Table \ref{tab:variable_definition} reports the eight baseline covariates. The underlying Compustat and ExecuComp fields are listed in Table \ref{tab:compustat_and_execucomp_variables} in Appendix \ref{app:data}.

\begin{table}[htb]
    \centering
    \caption{Empirical variable definitions}
    \small
    \begin{tabularx}{\textwidth}{
        l
        >{\hsize=.8\hsize\linewidth=\hsize\raggedright\arraybackslash}X
        >{\hsize=1.2\hsize\linewidth=\hsize\raggedright\arraybackslash}X
    }
        \hline\hline
        \textbf{Variable} & \textbf{Definition} & \textbf{In words} \\
        \hline
        Assets
        & $\log(AT_{j,t-1})$
        & Log of the parent firm's total assets in the previous year. \\
        \hline
        Leverage
        & $(DLC_{j,t-1}+DLTT_{j,t-1})/AT_{j,t-1}$
        & Short-term and long-term debt divided by total assets, all measured in the previous year. \\
        \hline
        Investment Intensity
        & $CAPX_{j,t-1}/PPENT_{j,t-2}$
        & Capital expenditure in the previous year relative to net property, plant, and equipment two years earlier. \\
        \hline
        Cash Holdings
        & $CHE_{j,t-1}/AT_{j,t-2}$
        & Cash and short-term investments in the previous year relative to total assets two years earlier. \\
        \hline
        Sales
        & $\log(SALE_{j,t})$
        & Log of the parent firm's sales in the current year. \\
        \hline
        Tobin Q
        & $(AT+CSHO\times PRCC\_F-SEQ-TXDITC)/AT$ in year $t$
        & Approximate market value of the parent firm relative to total assets in the current year. \\
        \hline
        CEO Age
        & CEO age in years
        & Age of the parent firm's CEO in the current year. \\
        \hline
        CEO Tenure
        & Year $t$ minus the year the executive became CEO
        & Number of years the executive has served as CEO by the current year. \\
        \hline
    \end{tabularx}
    \label{tab:variable_definition}
\end{table}

The variables should be interpreted as empirical proxies rather than direct measures of structural mechanisms. In particular, capital expenditure is not identified as environmental-purpose investment, and we do not claim a fix part as “green” investments, but broad capital expenditure, as the true amount put to environmental compliance is hard to observe without dedicated source. Several firm-level and aggregate variables may operate through more than one economic margin. Table \ref{tab:framework_channels} therefore links each channel to its framework notation and lists the empirical variables that can reasonably enter it. Repeated proxies across rows are deliberate: they allow, for example, assets, leverage, liquidity, and aggregate conditions to affect both production scale and the composite investment-cost wedge $\psi_t=\Psi(\mathcal{F}_t)$ without assigning a structural sign or a unique channel to any individual proxy.

\begin{table}[htb]
    \centering
    \caption{Framework channels, empirical proxies, and interpretive roles}
    \small
    \begin{tabularx}{\textwidth}{
        >{\raggedright\arraybackslash}p{0.24\textwidth}
        >{\raggedright\arraybackslash}p{0.28\textwidth}
        >{\raggedright\arraybackslash}X
    }
        \hline\hline
        \textbf{Framework channel} & \textbf{Empirical proxy} & \textbf{Interpretive role} \\
        \hline
        Scale and activity ($q_t$, $A_t$)
        & Sales, assets, Tobin's Q, cash holdings, leverage, FRED-QD principal components
        & Captures current operating scale, installed capacity, growth opportunities, and conditions that support or constrain production. \\
        \hline
        Investment-related adjustment ($I_t$, $G_t$, $\omega_t$, $m_{t+1}$)
        & Investment intensity, assets, Tobin's Q, cash holdings, leverage, CEO age, CEO tenure, FRED-QD principal components
        & Captures total investment, the share directed toward emissions-reducing adjustment, and the resulting change in future emissions intensity. \\
        \hline
        Investment-cost conditions ($\psi_t=\Psi(\mathcal{F}_t)$)
        & Assets, leverage, cash holdings, FRED-QD principal components
        & May shift the composite cost of environmentally oriented investment; the framework imposes no fixed functional form or structural sign for any individual proxy. \\
        \hline
        Managerial allocation ($\omega_t$, $\mathcal{F}_t$)
        & CEO age, CEO tenure
        & Captures heterogeneity in experience, persistence, and the allocation of investment between incumbent production and emissions-reducing adjustment. \\
        \hline
        Aggregate conditions ($A_t$, $\mathcal{F}_t$)
        & FRED-QD principal components
        & Summarizes common movements in demand, prices, credit, and other macroeconomic conditions that may affect both production scale and investment costs. \\
        \hline
    \end{tabularx}
    \label{tab:framework_channels}
\end{table}

For firm data completion (filling missing facility-year observations for facilities that were observed in at least one other year, and assigning zero emissions to these imputed observations, see Section \ref{sec:fac_data}), parent-firm identifiers are imputed separately: facility-years preceding a facility’s first TRI observation are assigned to the parent firm reported in that first observation, while years following its final observation retain the last reported parent firm. For gaps between observed years, the previously reported parent firm is carried forward until a subsequent TRI observation identifies a change in ownership, at which point the new parent firm is used. If the assigned parent firm did not yet exist or has no corresponding firm-level financial data in a particular year, this does not introduce artificial financial information because those fields remain missing; consequently, that facility will not enter the final balanced sample.

\subsubsection{Macroeconomic Data}    \label{sec:macro_data}

Aggregate conditions are measured using the March 2025 vintage of FRED-QD compiled by \citet{mccracken2020fred}. The quarterly series are first averaged to calendar years and then transformed using the published FRED-QD transformation codes shown in Equation \eqref{equ:tcode}. After retaining series with complete transformed observations throughout 1992--2023 and removing constants, the PCA input contains 237 annual series. Each series is standardized using its population standard deviation before principal components are estimated.

The first three components explain 33.92\%, 11.91\%, and 9.81\% of the variation in the transformed macroeconomic panel. The conditional facility models use the first component, merged many-to-one by calendar year so that every facility receives the same factor value in a given year. The aggregate-emissions analysis considers one-, two-, and three-component specifications. Because principal-component signs and loadings are statistical rather than structural, the components are treated as common macroeconomic factors rather than direct measures of scale or investment. Figure \ref{fig:pc1} displays the first-component score.

Formally, if $X_t=(x_{1t},\ldots,x_{pt})'$ is the vector of standardized transformed series, the first component is
\begin{equation}
    PC_{1,t}=a'_1X_t,
\end{equation}
where $a_1$ is the eigenvector associated with the largest eigenvalue of the covariance matrix and is normalized so that $a'_1a_1=1$.

\subsubsection{Summary Statistics}    \label{sec:summary_stats}

Table \ref{tab:summary_stats_emissions} presents facility-level emissions for the broad balanced-facility sample. The level measures contain 238,304 observations for 7,447 facilities over 1992--2023; 230,857 observations remain for symmetric annual changes in 1993--2023. The median is zero for every release category, reflecting both the concentration of emissions among a subset of facilities and the completed zero-release facility-years. Figure \ref{fig:agg_emissions} displays the amount of emissions by release channels.

\begin{table}[htb]
    \centering
    \caption{Summary statistics for facility emissions}
    \small
    \begin{tabular}{lrrrr}
        \hline\hline
        \textbf{Measure} & \textbf{N} & \textbf{Mean} & \textbf{Median} & \textbf{Std. Dev.} \\
        \hline
        Total Emissions (pounds)       & 238,304 & 27,630  & 0      & 305,753 \\
        Air Emissions (pounds)         & 238,304 & 20,186  & 0      & 149,423 \\
        Water Emissions (pounds)       & 238,304 & 485     & 0      & 12,194  \\
        Ground Emissions (pounds)      & 238,304 & 6,960   & 0      & 253,317 \\
        \%$\Delta$ Total Emissions     & 230,857 & -0.0302 & 0.0000 & 0.5551  \\
        \%$\Delta$ Air Emissions       & 230,857 & -0.0298 & 0.0000 & 0.5398  \\
        \%$\Delta$ Water Emissions     & 230,857 & -0.0020 & 0.0000 & 0.2933  \\
        \%$\Delta$ Ground Emissions    & 230,857 & -0.0020 & 0.0000 & 0.2273  \\
        \hline
    \end{tabular}
    \label{tab:summary_stats_emissions}
\end{table}

Parent-firm and CEO summaries in Table \ref{tab:summary_stats_firm_ceo} are calculated at the GVKEY-year level. Before variable-specific missingness, the broad sample contains 34,457 matched firm-years.

\begin{table}[htb]
    \centering
    \caption{Summary statistics for parent-firm financials and CEO characteristics}
    \small
    \begin{tabular}{lrrrr}
        \hline\hline
        \textbf{Measure} & \textbf{N} & \textbf{Mean} & \textbf{Median} & \textbf{Std. Dev.} \\
        \hline
        Assets (millions) & 21,805 & 10,818 & 1,405  & 39,560  \\
        Leverage                 & 21,612 & 0.2801 & 0.2557 & 0.2790  \\
        Investment Intensity     & 21,139 & 0.2744 & 0.1672 & 5.6471  \\
        Cash Holdings            & 21,318 & 0.1734 & 0.0612 & 7.7644  \\
        Sales (millions)         & 22,122 & 8,388  & 1,511  & 24,220  \\
        Tobin Q                  & 19,516 & 1.9198 & 1.4384 & 13.4036 \\
        CEO Age (years)          & 13,276 & 57.19  & 57.00  & 6.90    \\
        CEO Tenure (years)       & 12,993 & 6.71   & 5.00   & 7.09    \\
        \hline
    \end{tabular}
    \label{tab:summary_stats_firm_ceo}
\end{table}

These descriptive statistics use available observations in the broad balanced-facility file. The model samples additionally require 31-year completeness for the relevant covariate, which is why the estimation counts may be smaller and differ across variables.

\section{Results}    \label{sec:result}

\subsection{Baseline Analysis}

Guided by the economic framework presented earlier, we organize the results around the balance between scale and investment-related technique effects. Sales, assets, and Tobin's Q can capture differences in operating scale or growth opportunities, whereas investment intensity, cash holdings, and leverage affect firms' capacity to expand, replace capital, or finance emissions-reducing adjustments. Their coefficients are not structural scale or technique parameters: the sign and timing instead indicate which use of a variable is most visible in the data.

The baseline univariate TVMG analysis estimates each covariate on its own variable-specific balanced panel for 1993--2023. This preserves a complete 31-year history for every included facility while avoiding the large common-sample loss that would result from requiring every covariate to be observed simultaneously. Table \ref{tab:baseline_variable_specific_samples} reports the exact samples used in the baseline models.

\begin{table}[htb]
\centering
\caption{Baseline Variable-Specific Balanced Samples}
\begin{tabular}{lrrr}
\hline\hline
Variable & Facilities & Firms & Observations \\
\hline
Assets & 5,035 & 308 & 156,085 \\
Leverage & 5,171 & 335 & 160,301 \\
Investment Intensity & 4,841 & 302 & 150,071 \\
Cash Holdings & 5,028 & 320 & 155,868 \\
Sales & 5,106 & 316 & 158,286 \\
Tobin Q & 3,377 & 201 & 104,687 \\
CEO Age & 1,939 & 80 & 60,109 \\
CEO Tenure & 2,923 & 119 & 90,613 \\
\hline
\end{tabular}
\label{tab:baseline_variable_specific_samples}
\end{table}

We first estimate the relationship between facility emissions growth and each of eight baseline covariates---assets, leverage, investment intensity, cash holdings, sales, Tobin's Q, CEO age, and CEO tenure---using
\begin{equation}
    \%\Delta Emission_{it} = \alpha_{it} + \beta_{t}Var_{J(i,t)} + u_{it},
    \label{equ:baseline}
\end{equation}
where the outcome is the symmetric emissions-growth measure in Equation \eqref{equ:emissionchange}. The estimation uses a Gaussian kernel, the facility as the mean-group unit, and the mean-group aggregation in Equation \eqref{equ:mg}.

\begin{figure}[htb]
    \centering
    \includegraphics[width=0.95\textwidth]{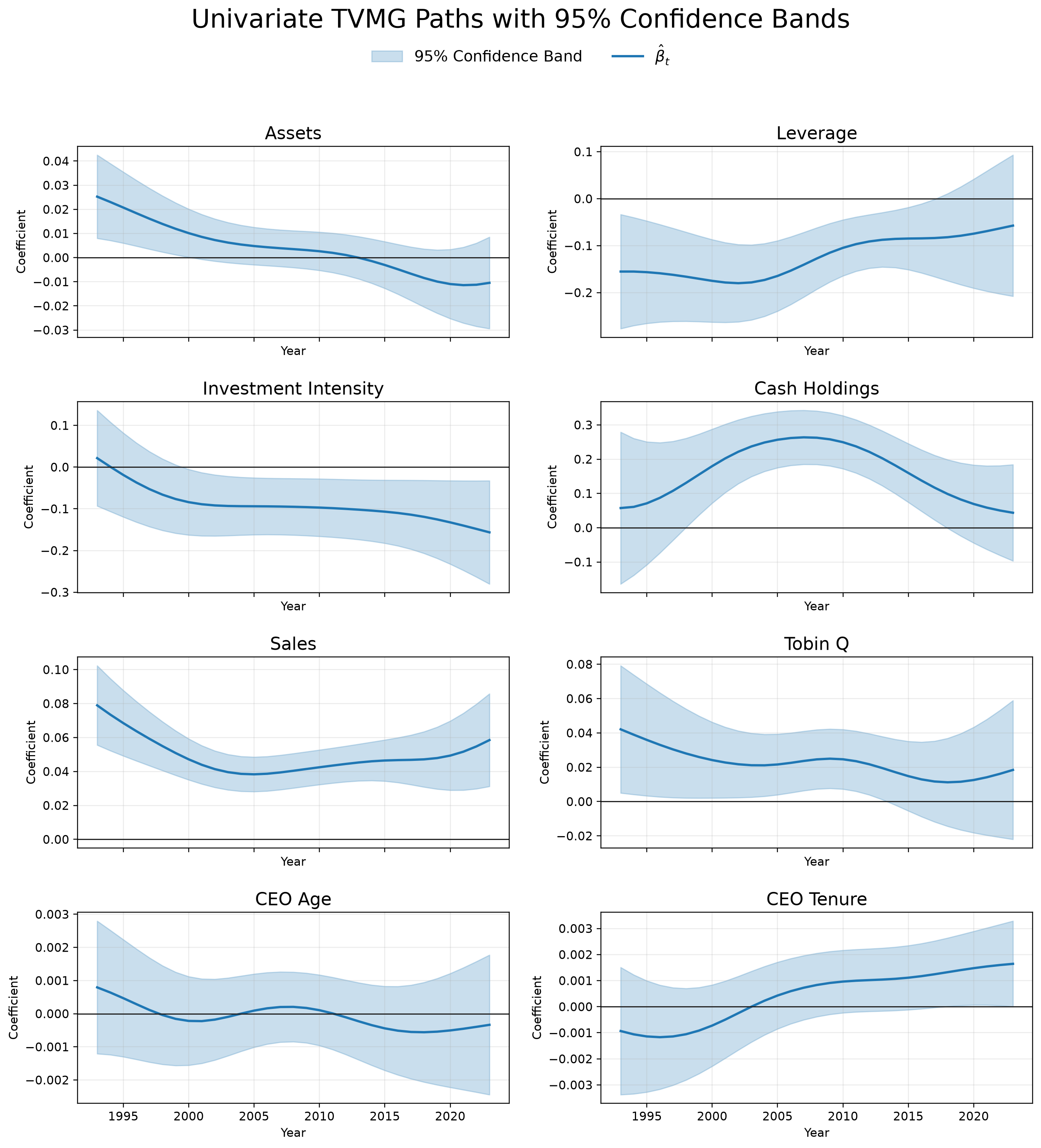}
    \caption{Baseline univariate TVMG coefficient paths for the eight variable-specific balanced samples. Shaded regions are 95\% confidence bands.}
    \label{fig:baseline_pointwise_paths}
\end{figure}

The estimates use the fixed bandwidth $H=\sqrt{31}\approx5.57$ and 95\% confidence bands. The fixed square-root rule is retained for a common smoothing convention across variables and specifications. Table \ref{tab:variable_significant_periods} reports the exact significant periods in the baseline estimates, and Figure \ref{fig:baseline_pointwise_paths} displays the corresponding coefficient paths and confidence bands.

\begin{table}[htb]
    \centering
    \caption{Baseline TVMG model results}
    \begin{tabular}{lcc}
        \hline\hline
        \textbf{Variable}    & \textbf{Direction} & \textbf{Significant Period(s)} \\
        \hline
        Assets                & Positive & 1993--2000 \\
        Leverage              & Negative & 1993--2017 \\
        Investment Intensity  & Negative & 2000--2023 \\
        Cash Holdings         & Positive & 1998--2017 \\
        Sales                 & Positive & 1993--2023 \\
        Tobin Q               & Positive & 1993--2013 \\
        CEO Age               & ---      & None       \\
        CEO Tenure            & Positive & 2018--2023 \\
        \hline
    \end{tabular}
    \label{tab:variable_significant_periods}
\end{table}

The clearest scale-related result is the positive sales coefficient throughout 1993--2023. Higher parent-firm sales remain associated with faster facility emissions growth across the entire sample, consistent with a persistent relationship between operating scale and release growth even as regulation and production technologies change. Positive coefficients on assets in 1993--2000 and Tobin's Q in 1993--2013 reinforce this interpretation: larger asset bases and stronger growth opportunities coincide with higher emissions growth when they are examined separately. This pattern is consistent with the general link between economic activity and environmental pressure \citep{hertwich2010assessing,heutel2012should}.

Investment intensity provides the main countervailing pattern. Its coefficient is negative from 2000 through 2023, consistent with observed capital expenditure becoming associated on net with capital replacement, process improvement, fuel substitution, or abatement rather than only capacity expansion. Because CapEx/PPE is not a measure of environmental-purpose investment, this result should be read as an investment-related technique association, not as a direct estimate of green investment. The long window is nevertheless consistent with evidence that the decline in U.S. manufacturing pollution has been driven substantially by reductions in emissions intensity \citep{shapiro2018pollution}; disclosure and regulatory developments during the later part of the window may have strengthened the return to cleaner adjustment \citep{yang2021real,tomar2023greenhouse}.

The financial variables clarify how funding conditions can alter the balance between investment and scale. Cash holdings are positively associated with emissions growth from 1998--2017. In the data, liquidity therefore appears to accompany the ability to sustain production or undertake expansion more strongly than it proxies emissions-reducing investment. Leverage is negative from 1993--2017, which is consistent with debt burdens restraining capacity growth and operational expansion. Together, these signs show why finance variables cannot be assigned to pre-specified channel mechanically: cash can fund either capacity or cleaner capital, while leverage can restrict both, and the observed coefficient records their net association with emissions growth.

The managerial results are more limited. CEO age is never statistically significant, whereas CEO tenure is positive only in 2018--2023. The late tenure association is consistent with persistent managerial and organizational choices affecting how firms allocate investment between maintaining incumbent production and undertaking cleaner adjustment, but the result does not support a broad age-based managerial effect. The wider literature motivates managerial heterogeneity as a relevant channel \citep{manner2010impact,huang2013impact}; here, the evidence is specific to tenure and the end of the sample.

Taken together, the univariate results show simultaneous but opposing margins. Sales, early asset size, Tobin's Q, and cash holdings load on the expansion side, whereas investment intensity and leverage are negatively associated with emissions growth. The persistence of positive sales alongside negative investment intensity is particularly informative: firms can expand activity while capital spending is associated with lower emissions growth, so scale and technique forces need not move in the same direction. We next examine whether relocation exposure accounts for this baseline pattern, and investigate sectoral heterogeneity. Additional robustness sections follow. Subsequent sections examine variables, especially scale vs. investment simultaneously with joint estimations.

\FloatBarrier

\paragraph{Potential relocation exposure}

A potential interpretation concern is that lower domestic emissions may reflect foreign production relocation rather than investment in cleaner domestic production. We therefore interact each baseline variable with two firm-year proxies for multinational exposure: an indicator for positive foreign pretax income, $1(PIFO_{jt}>0)$, and non-negative foreign pretax income scaled by assets, $\max(PIFO_{jt},0)/AT_{j,t-1}$. The 95\% confidence band results contain only two interaction windows: the positive-PIFO indicator moderates assets positively in 2023, and the scaled measure moderates CEO tenure positively in 2000--2011. Neither window overlaps the corresponding baseline significance period for assets (1993--2000) or CEO tenure (2018--2023); the interactions for sales, investment intensity, cash holdings, leverage, Tobin's Q, and CEO age are not significant. Thus, the foreign-exposure exercise does not explain the baseline patterns. Table \ref{tab:relocation} in Appendix \ref{app:relocation} reports the full results.

\paragraph{Sectoral analysis}

To examine whether the baseline heterogeneity is organized by production setting, we repeat the eight univariate specifications within three-digit facility NAICS industries. Appendix \ref{app:sector} details the setup. Facility classification gives a clearer account of the balance between scale and investment-related adjustment. Sales has a positive significant period in 13 of the 17 industries, although furniture manufacturing also has an earlier negative period. Investment intensity is significant in five industries. It is negative in wood products (321), nonmetallic minerals (327), machinery (333), and miscellaneous manufacturing (339), but positive in printing and related support activities (323). In each of the four industries with a negative investment-intensity association, sales is positive over at least one period. This combination is consistent with operating scale raising emissions while capital expenditure is associated with capital replacement, process improvement, or other reductions in emissions intensity. The positive investment-intensity result in printing shows that capital spending can instead coincide with expansion. Heterogeneity across industries is expected, based on their natures, like machinery can benefit more from invested capital to the extent of toxic release control, while investments in sectors like printing are more about growth opportunities.

\FloatBarrier

\subsection{Robustness Checks}

We assess the baseline TVMG results along five main dimensions. First, simultaneous bands provide a more conservative account of uncertainty across the full coefficient path. Second, the models are re-estimated after removing synthetic-zero rows. Third, the outcome is aggregated to the parent-firm level. Fourth, facilities associated with the five largest parent-firm portfolios are excluded. Fifth, the bandwidth and kernel are varied.

\subsubsection{Uniform Bands for the Eight Baseline Paths}

The confidence bands in Figure \ref{fig:baseline_pointwise_paths} assess uncertainty one year at a time. Reading significance from 31 separate annual intervals may nevertheless overstate the evidence because it does not account for repeated comparisons across the path. We therefore complement them with 95\% simultaneous bands that cover the entire 1993--2023 coefficient path. These bands are deliberately more conservative than the pointwise intervals: a coefficient is reported as different from zero only when the simultaneous band excludes zero in that year.

Let $\widehat\beta_{it}$ be the local time-varying slope for facility $i$ in year $t$, $\widehat\beta_t=N^{-1}\sum_i\widehat\beta_{it}$, and $\widehat\sigma_t^2=N^{-1}\sum_i(\widehat\beta_{it}-\widehat\beta_t)^2$, so that $\widehat{SE}(\widehat\beta_t)=\widehat\sigma_t/\sqrt{N}$. For multiplier draw $b$, let $\xi_i^{(b)}$ be an independent Rademacher variable taking values $-1$ and $1$ with equal probability. The standardized multiplier process, simultaneous critical value, and band are
\begin{equation}
    \begin{aligned}
    G^{*(b)}(t)
    &=\frac{1}{\sqrt{N}}\sum_{i=1}^{N}
      \xi_i^{(b)}
      \frac{\widehat\beta_{it}-\widehat\beta_t}{\widehat\sigma_t}, \\
    c_{0.95}^{\mathrm{sup}}
    &=Q_{0.95}\!\left(
      \max_{t\in\mathcal{T}}\left|G^{*(b)}(t)\right|
      \right), \\
    \widehat{\mathcal{B}}_{0.95}(t)
    &=\left[
      \widehat\beta_t-c_{0.95}^{\mathrm{sup}}\widehat{SE}(\widehat\beta_t),
      \widehat\beta_t+c_{0.95}^{\mathrm{sup}}\widehat{SE}(\widehat\beta_t)
      \right],
      \qquad t\in\mathcal{T}.
    \end{aligned}
    \label{equ:baseline_uniform_band}
\end{equation}
Here $Q_{0.95}$ is the empirical 95th percentile across 2,000 multiplier draws and $\mathcal{T}=\{1993,\ldots,2023\}$. A common multiplier $\xi_i^{(b)}$ is used for all years of a facility, preserving the across-year covariance induced by kernel smoothing. Because the critical value is based on the largest absolute standardized fluctuation over the full path, it is generally larger than the pointwise normal critical value. The adjustment is made over time within each variable-specific path; it does not combine the eight different samples into one common family.

\begin{figure}[htbp]
    \centering
    \includegraphics[width=0.95\textwidth]{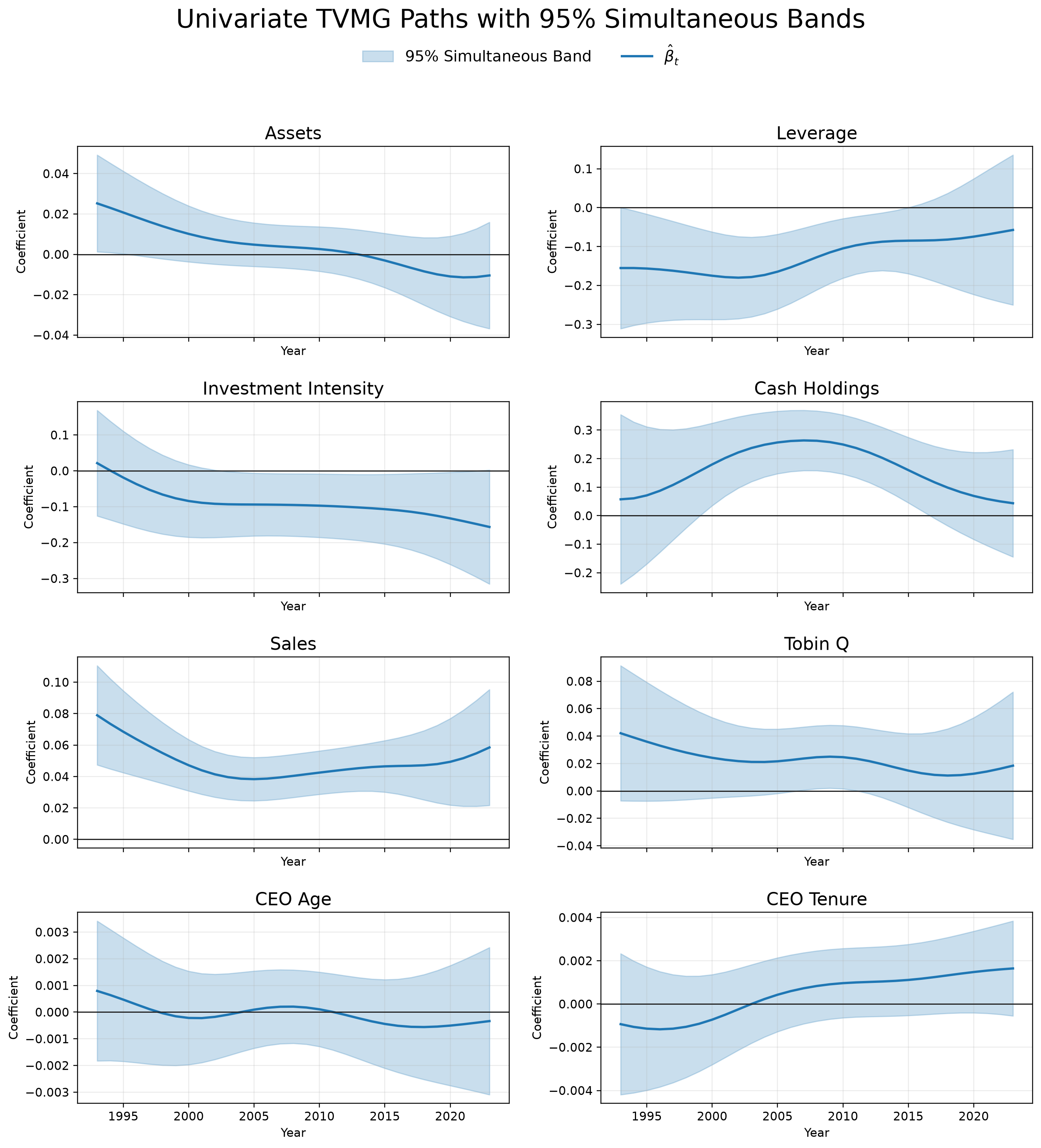}
    \caption{Baseline univariate TVMG coefficient paths with 95\% simultaneous (uniform) bands.}
    \label{fig:baseline_uniform_paths}
\end{figure}

Figure \ref{fig:baseline_uniform_paths} displays the eight coefficient paths with the conservative full-grid bands, reported in Table \ref{tab:baseline_uniform_bands}. The more conservative bands preserve the main economic contrast while shortening several windows. Sales remains positive throughout 1993--2023. Investment intensity remains negative in 2003--2022, leverage in 1994--2014, and cash holdings is positive in 2000--2016. The assets window contracts to 1993--1995 and the Tobin's Q window to 2007--2011, while the bands for CEO age and CEO tenure never exclude zero. These results show that the main scale--investment pattern, as well as many other associations are not an artifact of inspecting separate annual pointwise intervals, while the shorter windows also provide a more cautious account of its timing.

\begin{table}[htb]
    \centering
    \caption{Conservative simultaneous bands for the eight baseline coefficient paths}
    \small
    \begin{tabular}{lcc}
        \hline\hline
        \textbf{Variable} & \textbf{Direction} & \textbf{Significant Period(s)} \\
        \hline
        Assets               & Positive & 1993--1995 \\
        Leverage             & Negative & 1994--2014 \\
        Investment Intensity & Negative & 2003--2022 \\
        Cash Holdings        & Positive & 2000--2016 \\
        Sales                & Positive & 1993--2023 \\
        Tobin's Q            & Positive & 2007--2011 \\
        CEO Age              & ---      & None       \\
        CEO Tenure           & ---      & None       \\
        \hline
    \end{tabular}
    \begin{minipage}{0.94\textwidth}
    \textit{Notes:} The table reports years in which the 95\% simultaneous band for the full 1993--2023 path excludes zero. The bands use 2,000 facility-level Rademacher multiplier draws and control coverage over time separately for each variable-specific sample.
    \end{minipage}
    \label{tab:baseline_uniform_bands}
\end{table}

\FloatBarrier

\subsubsection{Excluding Synthetic-Zero Rows}

The completed panel assigns a zero contribution to observed TRI releases when a facility-year is inserted to bridge a reporting gap. This accounting convention is useful for representing entry and exit in a balanced design (non-reporting may exist but low amount and impact overall, discussion in Section \ref{sec:fac_data}). We would like to know the degree of contribution of facility establishment and shut-down, and more broadly, data-completion, to our tests. We therefore re-estimate the eight one-regressor models after removing every inserted facility-year while retaining observations in which the facility genuinely reported zero releases.

\begin{figure}[htbp]
    \centering
    \includegraphics[width=0.95\textwidth]{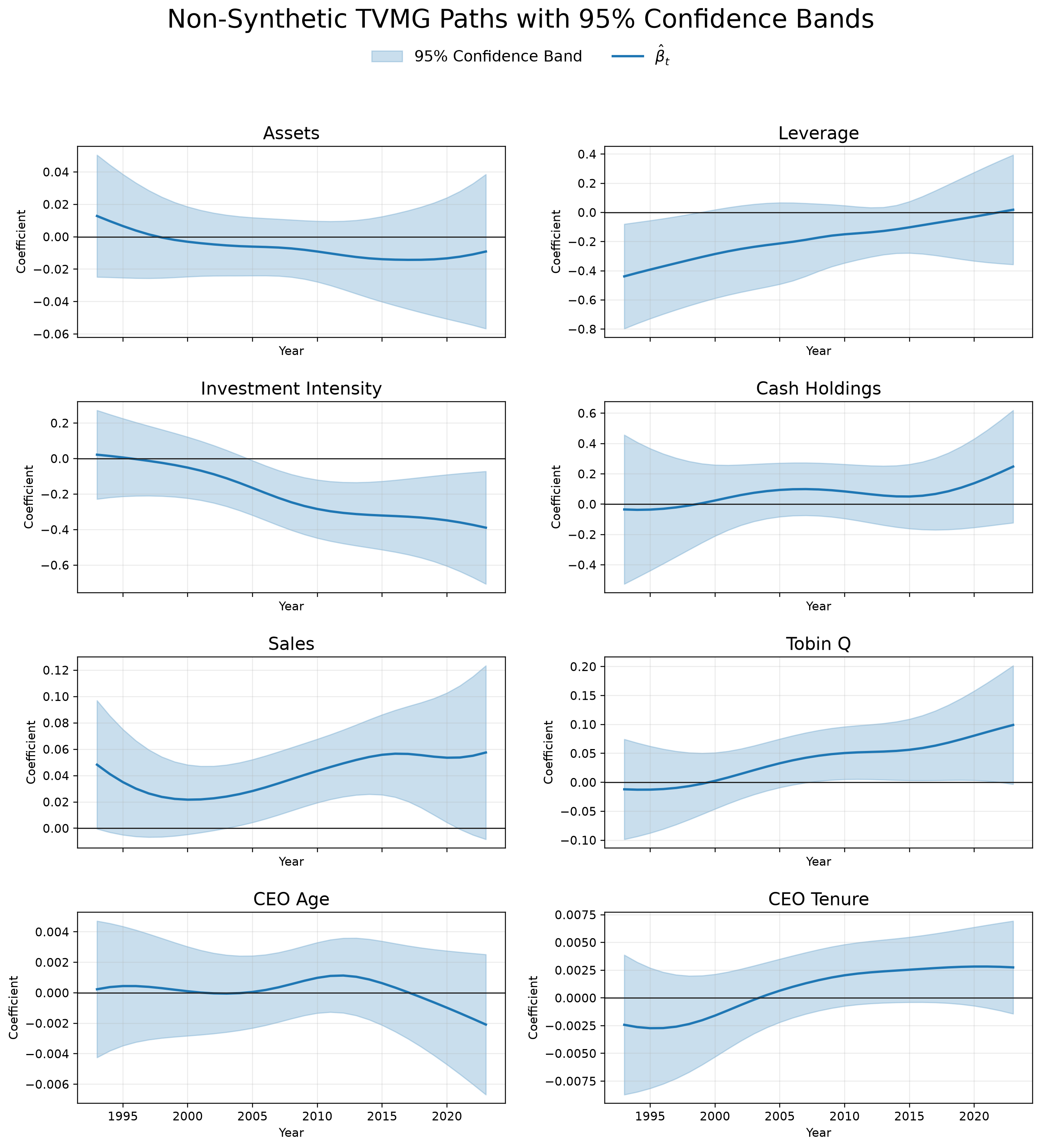}
    \caption{Univariate TVMG coefficient paths after excluding synthetic-zero rows. Shaded regions are 95\% confidence bands.}
    \label{fig:baseline_non_synthetic_paths}
\end{figure}

The strict source contains 738,590 facility-year rows from 57,099 facilities, no residual imputed rows, and 189,214 genuinely reported total-zero observations. For each regressor, we again require the outcome and the selected covariate to be not-nan in every year from 1993 through 2023. Table \ref{tab:baseline_non_synthetic} reports the significant periods and sample sizes, while Figure \ref{fig:baseline_non_synthetic_paths} displays the full coefficient paths.

\begin{table}[htb]
    \centering
    \caption{Baseline sensitivity excluding synthetic-zero rows}
    \small
    \begin{tabular}{lccr}
        \hline\hline
        \textbf{Variable} & \textbf{Direction} & \textbf{Significant Period(s)} &
        \textbf{Facilities} \\
        \hline
        Assets               & ---      & None       & 1,100 \\
        Leverage             & Negative & 1993--1998 & 1,091 \\
        Investment Intensity & Negative & 2005--2023 & 1,006 \\
        Cash Holdings        & ---      & None       & 1,046 \\
        Sales                & Positive & 2003--2020 & 1,109 \\
        Tobin's Q            & Positive & 2008--2021 &   731 \\
        CEO Age              & ---      & None       &   445 \\
        CEO Tenure           & ---      & None       &   628 \\
        \hline
    \end{tabular}
    \begin{minipage}{0.96\textwidth}
    \footnotesize
    \end{minipage}
    \label{tab:baseline_non_synthetic}
\end{table}

The central scale--investment contrast remains visible without data-completion. Investment intensity is negative in 2005--2023, while sales is positive in 2003--2020. Tobin's Q is positive in 2008--2021, and leverage is negative only in 1993--1998. Assets, cash holdings, CEO age, and CEO tenure have no significant period. Thus, the long negative investment association and positive sales association are not generated solely by the inserted zero-release observations. The absence of asset and cash-holdings periods in the reporting sample indicates that confidence in those relationships is sensitive to the panel-construction rule, potentially depend on opening and shutting facilities. For example, if excess cash is assumed to be used for expansion primarily through the establishment of new facilities, the disappearance of the positive cash coefficient after excluding facility openings is expected, as the expansion channel is effectively removed.

\FloatBarrier

\subsubsection{Firm-Level Alternative}

The main analysis retains the facility as the estimation unit. TRI reporting coverage, the TRIFID reporting identifier, and NAICS-based production classification are determined at the facility level. Aggregating across a parent firm's facilities can therefore combine establishments with different technologies, industries, reporting histories, and emissions trajectories. The facility-level design preserves this production-level heterogeneity and makes the inclusion and exclusion rules correspond to the unit at which TRI emissions are reported. We consequently focus the main discussion on the facility-level estimates.

\begin{figure}[htbp]
    \centering
    \includegraphics[width=0.95\textwidth]{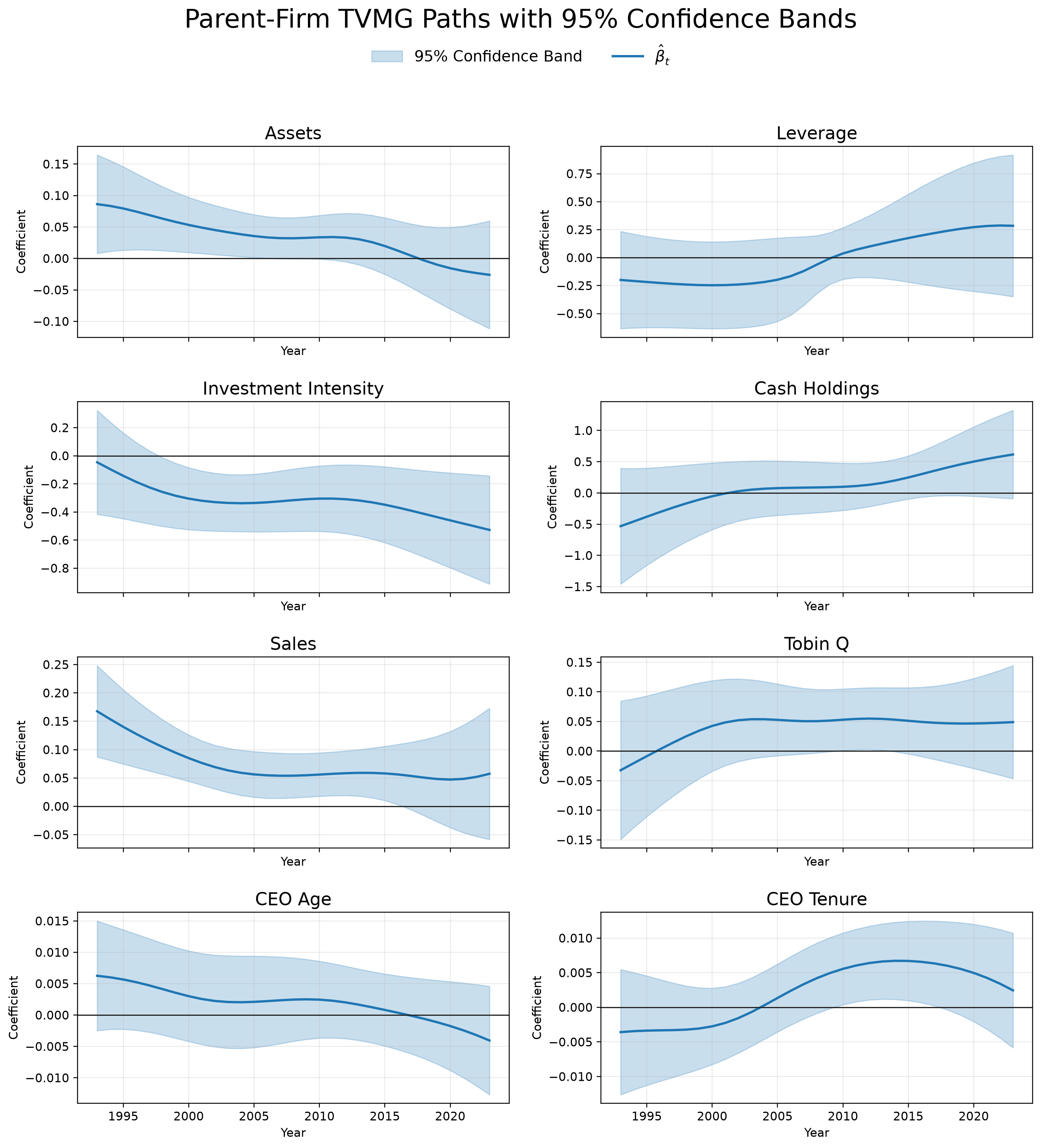}
    \caption{Parent-firm level univariate TVMG coefficient paths. Shaded regions are 95\% confidence bands.}
    \label{fig:firm_level_alternative_paths}
\end{figure}

As an alternative, we aggregate the outcome to the firm level as a supplement test. For each variable-specific balanced sample, a facility must have a non-missing and unchanged parent GVKEY in every year from 1992--2023. We then sum current and lagged facility emissions within each GVKEY-year and recompute the symmetric emissions-growth outcome from those firm totals. Table \ref{tab:firm_level_alternative} summarizes the comparison, and Figure \ref{fig:firm_level_alternative_paths} displays the full parent-firm coefficient paths.

\begin{table}[htbp]
    \centering
    \caption{Facility- and firm-level univariate TVMG results}
    \small
    \begin{tabularx}{\textwidth}{lXXrr}
        \hline\hline
        \textbf{Variable} & \textbf{Facility-Level Baseline} &
        \textbf{Firm-Level Alternative} & \textbf{Firms} &
        \textbf{Firm-Years} \\
        \hline
        Assets                & 1993--2000 (+)   & 1993--2006 (+)   & 308 & 9,548  \\
        Leverage              & 1993--2017 ($-$) & None              & 335 & 10,385 \\
        Investment Intensity  & 2000--2023 ($-$) & 1998--2023 ($-$) & 302 & 9,362  \\
        Cash Holdings         & 1998--2017 (+)   & None              & 320 & 9,920  \\
        Sales                 & 1993--2023 (+)   & 1993--2016 (+)   & 313 & 9,703  \\
        Tobin Q               & 1993--2013 (+)   & 2010--2013 (+)   & 199 & 6,169  \\
        CEO Age               & None             & None              & 80  & 2,480  \\
        CEO Tenure            & 2018--2023 (+)   & 2010--2017 (+)   & 117 & 3,627  \\
        \hline
    \end{tabularx}
    \begin{minipage}{0.96\textwidth}
    \footnotesize
    \textit{Notes:} A plus sign in parentheses denotes a positive coefficient, and a minus sign denotes a negative coefficient.
    \end{minipage}
    \label{tab:firm_level_alternative}
\end{table}

The firm-level alternative preserves six of the eight qualitative baseline patterns, although it does not reproduce every endpoint. Assets, investment intensity, sales, and Tobin's Q retain their baseline directions, CEO age remains insignificant, and CEO tenure remains positive but shifts to an earlier period. The negative leverage and positive cash-holdings coefficients are no longer significant after firm aggregation. Thus, the central scale--investment contrast survives: sales is positive through 2016, while investment intensity is negative from 1998--2023.

The following influential-firm exclusion exercise in Section \ref{sec:five_firm} provides the complementary five-firm-out check. It removes the five parent firms associated with the largest facility portfolios while retaining the facility as the estimation unit, thereby testing whether the main facility-level results are driven by a small number of highly represented firms.

\FloatBarrier

\subsubsection{Excluding the Five Firms with the Most Facilities}    \label{sec:five_firm}

To test whether a small group of firms' facilities dominated the results, for each variable-specific balanced sample, we rank parent firms by the number of distinct facilities ever associated with them, break ties by GVKEY, and remove the five highest-ranked firms in one joint exclusion. Any facility ever associated with one of these firms is removed in every year, after which the facility-level TVMG model is re-estimated once. This design directly tests whether the results reflect firms with unusually large facility portfolios.

We compare the exclusion estimate $\hat{\beta}^{(-5)}_{MG,t}$ with the full-sample estimate using
\begin{equation*}
    D_t=\frac{\left|\hat{\beta}^{(-5)}_{MG,t}-\hat{\beta}_{MG,t}\right|}
    {SE(\hat{\beta}_{MG,t})}
\end{equation*}
and record whether the coefficient changes sign. Table \ref{tab:top_five_firm_exclusion} reports the maximum $D_t$ within each variable's baseline significant window. The exclusion removes 730 facilities for the six financial variables, 671 for CEO age, and 682 for CEO tenure because the variable-specific samples differ.

\begin{table}[htbp]
    \centering
    \caption{Robustness to excluding the five firms with the most facilities}
    \small
    \begin{tabularx}{\textwidth}{lrrX}
        \hline\hline
        \textbf{Variable} & \textbf{Facilities Removed} &
        \textbf{Max. $D_t$ in Baseline Window} & \textbf{Sign Flip in Window} \\
        \hline
        Assets               & 730 & 0.13 & No \\
        Leverage             & 730 & 1.51 & No \\
        Investment Intensity & 730 & 0.67 & No \\
        Cash Holdings        & 730 & 1.55 & No \\
        Sales                & 730 & 0.81 & No \\
        Tobin Q              & 730 & 0.52 & No \\
        CEO Age              & 671 & ---  & Not applicable \\
        CEO Tenure           & 682 & 0.24 & No \\
        \hline
    \end{tabularx}
    \label{tab:top_five_firm_exclusion}
\end{table}

No coefficient changes sign within its baseline significant period. The direction of the results is therefore not generated by the five largest facility portfolios. Magnitudes are more sensitive for leverage and cash holdings, whose maximum deviations are about 1.5 full-sample standard errors; the remaining significant-window deviations are below one standard error.

\subsubsection{Smoothness Sensitivity}    \label{sec:sensitivity}

\begin{figure}[htbp]
    \centering
    \includegraphics[width=0.95\textwidth]{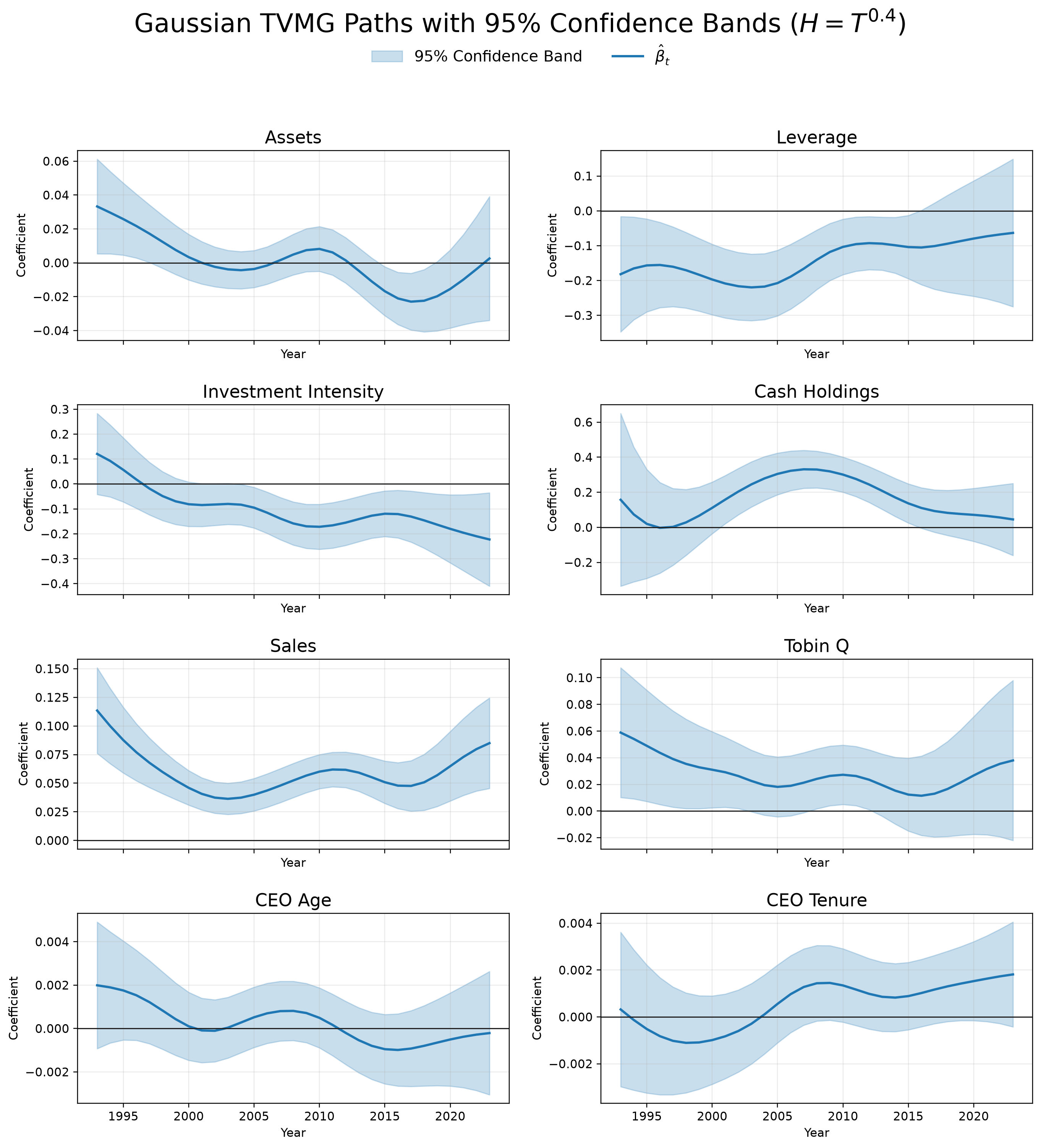}
    \caption{Univariate TVMG sensitivity paths using the Gaussian kernel and bandwidth $H=T^{0.4}$. Shaded regions are 95\% confidence bands.}
    \label{fig:sensitivity_bandwidth_t04_paths}
\end{figure}

\begin{figure}[htbp]
    \centering
    \includegraphics[width=0.95\textwidth]{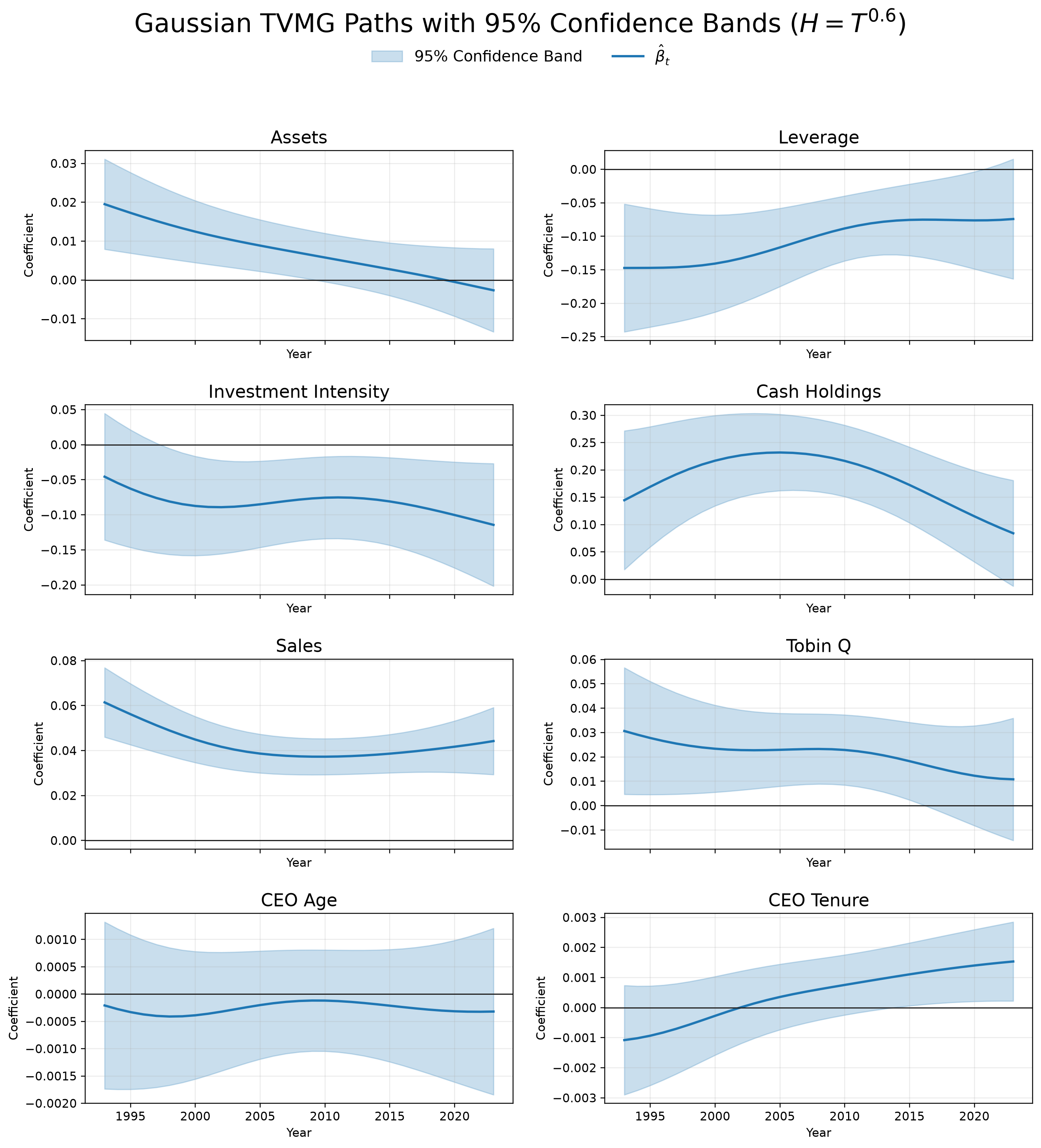}
    \caption{Univariate TVMG sensitivity paths using the Gaussian kernel and bandwidth $H=T^{0.6}$. Shaded regions are 95\% confidence bands.}
    \label{fig:sensitivity_bandwidth_t06_paths}
\end{figure}

Time-varying models depend on smoothness and bandwidth settings. To investigate, we re-estimate all eight baseline specifications with Gaussian bandwidths $H=T^{0.4}$ and $H=T^{0.6}$ and with an Epanechnikov kernel at $H=T^{0.5}$. Figures \ref{fig:sensitivity_bandwidth_t04_paths} and \ref{fig:sensitivity_bandwidth_t06_paths} display the full paths under the two alternative Gaussian bandwidths. A smaller bandwidth naturally yields more variation, while a larger bandwidth produces smoother paths. The alternative Gaussian bandwidths preserve the main directions while changing the width of the significant windows. Sales remains positive throughout 1993--2023 under both bandwidths. Investment intensity is negative in 2004--2023 with $T^{0.4}$ and 1998--2023 with $T^{0.6}$; leverage is negative in 1993--2015 and 1993--2020, respectively. Cash holdings remains positive, although its window ranges from 2001--2015 under the narrower bandwidth to 1993--2022 under the wider bandwidth. The asset and Tobin's Q paths are more timing-sensitive, but their early positive scale associations remain visible.

For an alternative kernel function choice, we re-estimate the models using the Epanechnikov kernel in place of the Gaussian kernel, with
\begin{equation*}
    K(u)
    =
    \begin{cases}
        \frac{3}{4}(1-u^{2}), & |u| \le 1, \\
        0, & |u| > 1.
    \end{cases}
\end{equation*}

\begin{figure}[htbp]
    \centering
    \includegraphics[width=0.95\textwidth]{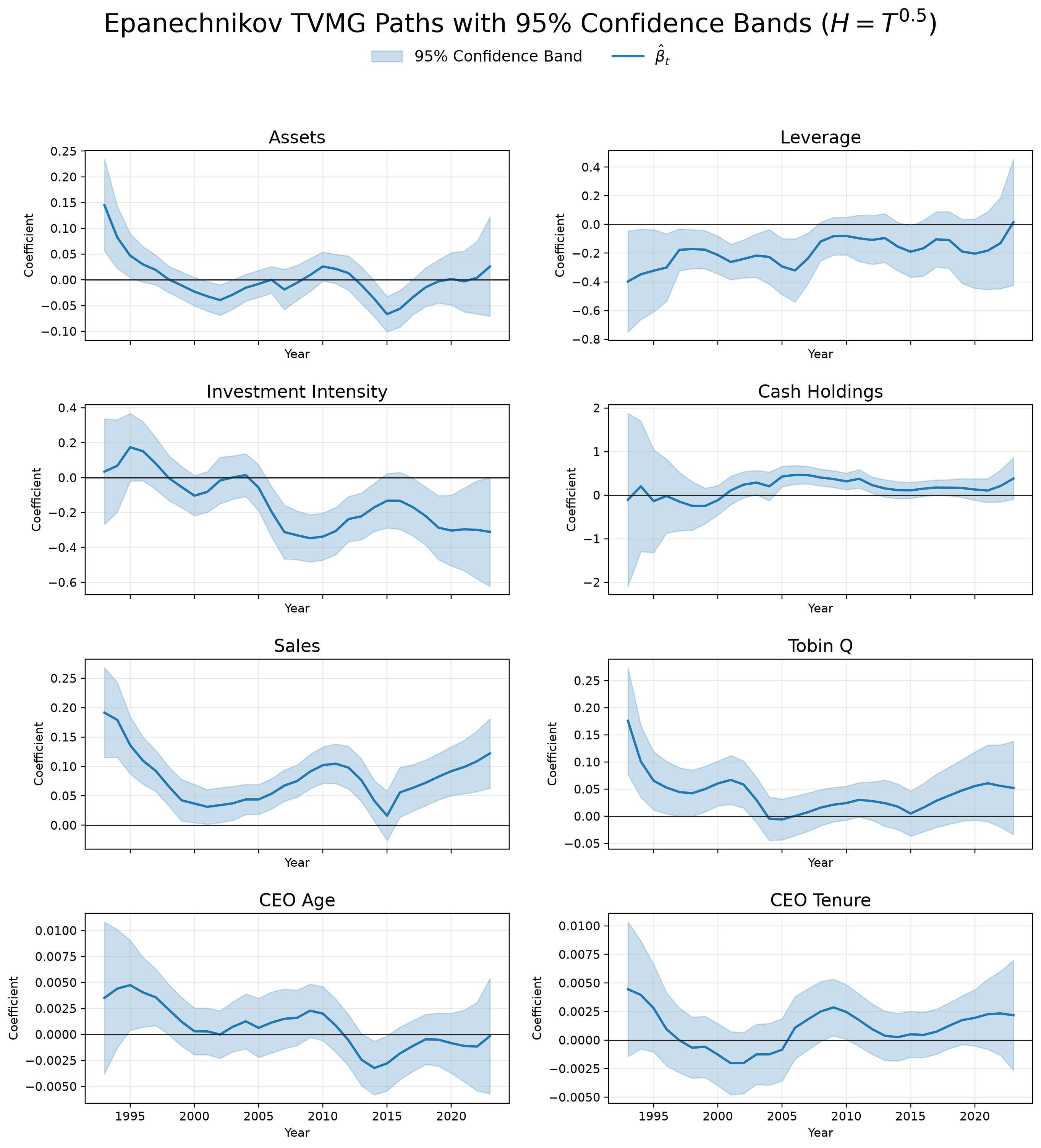}
    \caption{Univariate TVMG sensitivity paths using the Epanechnikov kernel and bandwidth $H=T^{0.5}$. Shaded regions are 95\% confidence bands.}
    \label{fig:sensitivity_ep_kernel_paths}
\end{figure}

Figure \ref{fig:sensitivity_ep_kernel_paths} displays the coefficient paths under the Epanechnikov kernel. The compact-support kernel produces shorter and sometimes split periods, as expected when distant years receive zero weight. Even so, sales remains positive in 1993--2014 and 2016--2023, investment intensity remains negative in 2006--2014 and 2017--2022, and leverage remains predominantly negative. The isolated CEO-age and CEO-tenure periods under this kernel confirm that the managerial paths are less robust than the financial and scale paths. Overall, the central results does not depend on a single reasonable smoothing choice, although exact endpoints should not be interpreted as structural break dates.

\FloatBarrier

\subsection{Conditional Relationships under Aggregate Economic Conditions}

The baseline coefficients can combine firm-specific scale or investment choices with common movements in demand and financing conditions. We therefore pair each of the eight variables with the annual first principal component of FRED-QD. Because the factor is common to all facilities in a year, this two-regressor specification asks whether the firm variable retains an association after common macroeconomic variation is held constant.

With the additional regressor, the model becomes
\begin{equation}
    \%\Delta Emission_{it} = \alpha_{it} + \beta_{1,t}Var_{J(i,t)} + \beta_{2,t}PC_t + u_{it},
\end{equation}
where $PC_t$ is the first principal component estimated from FRED-QD. Each model uses the same variable-specific balanced sample as its baseline counterpart in Table \ref{tab:baseline_variable_specific_samples}.

\begin{figure}[htbp]
    \centering
    \includegraphics[width=0.95\textwidth]{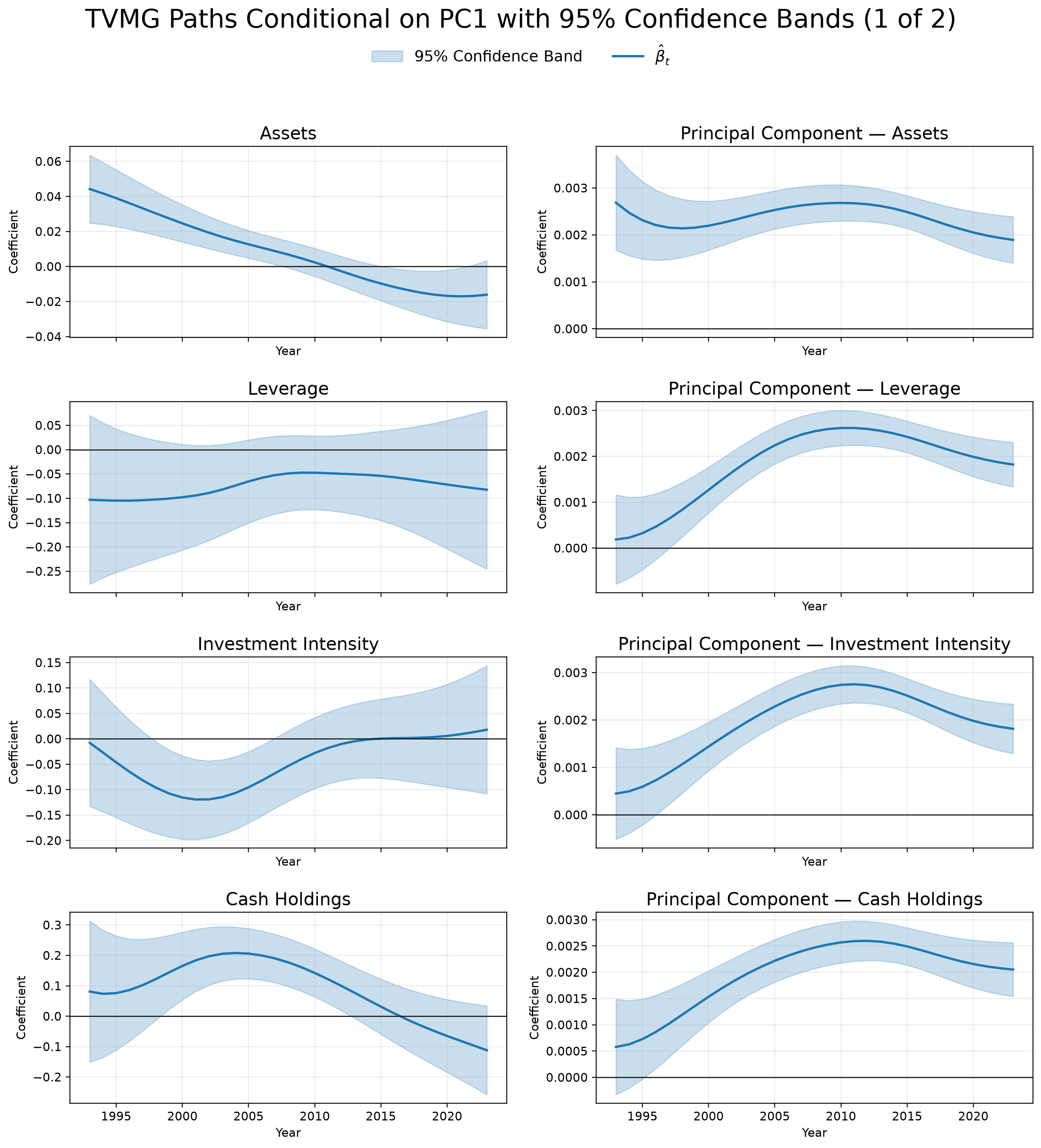}
    \caption{Two-regressor TVMG coefficient paths conditional on the first principal component of FRED-QD for the first set of firm variables. Shaded regions are 95\% confidence bands.}
    \label{fig:with_pc_1}
\end{figure}

\begin{figure}[htbp]
    \centering
    \includegraphics[width=0.95\textwidth]{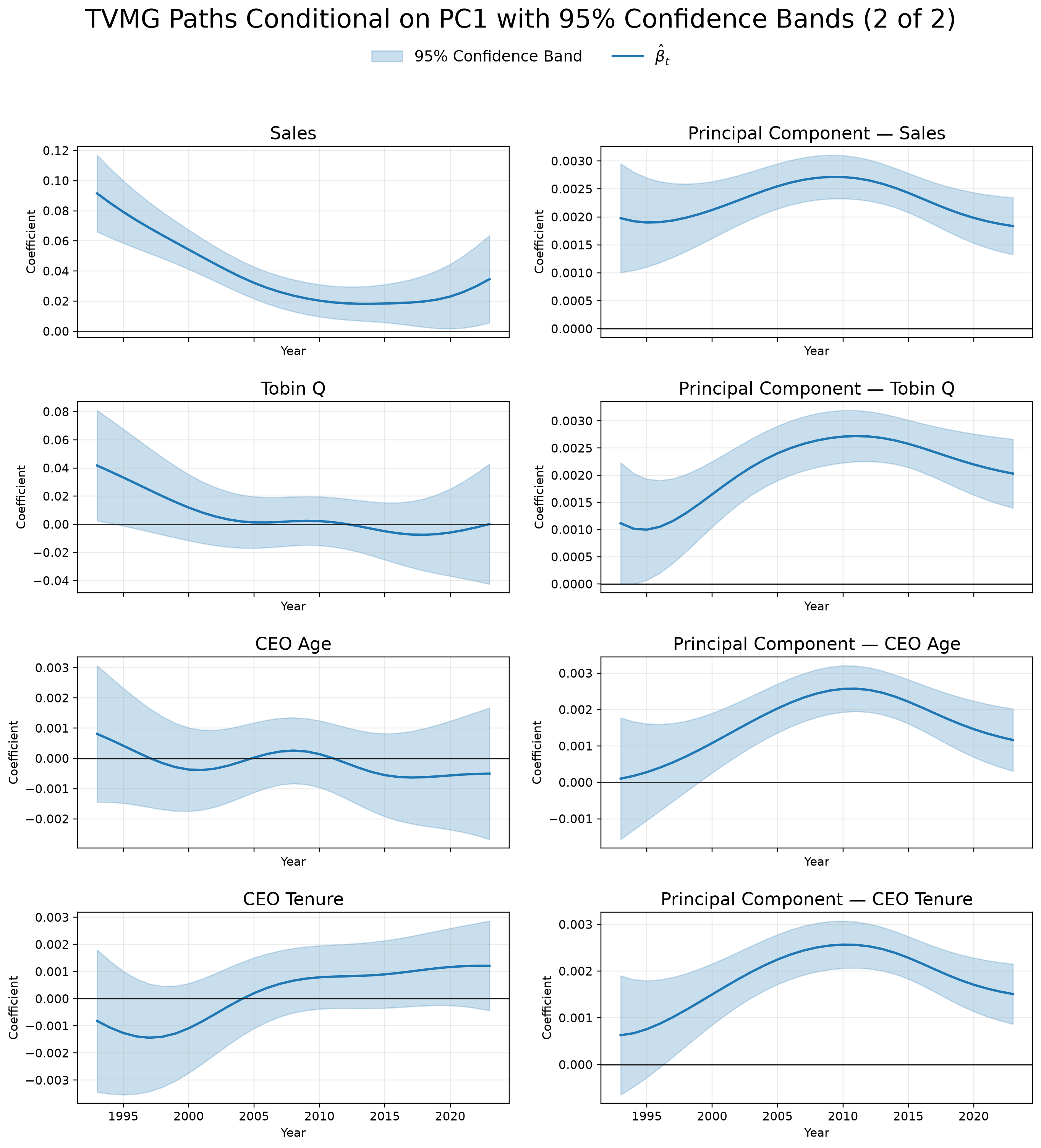}
    \caption{Two-regressor TVMG coefficient paths conditional on the first principal component of FRED-QD for the remaining firm variables. Shaded regions are 95\% confidence bands.}
    \label{fig:with_pc_2}
\end{figure}

Figures \ref{fig:with_pc_1} and \ref{fig:with_pc_2} display the paths, while Table \ref{tab:variable_significant_periods_with_pc} reports both the firm-variable coefficient and the PC1 coefficient from each model. Macro conditioning varies the baseline windows: leverage and both CEO variables lose significance, the investment and Tobin's Q windows contract, and the asset coefficient reverses sign late in the sample.

\begin{table}[htb]
    \centering
    \caption{Two-regressor TVMG results conditional on the FRED-QD first principal component}
    \small
    \begin{tabularx}{\textwidth}{lXX}
        \hline\hline
        \textbf{Firm Variable} & \textbf{Firm-Variable Significant Period(s)} &
        \textbf{PC1 Significant Period(s)} \\
        \hline
        Assets               & 1993--2007 (+); 2016--2021 ($-$) & 1993--2023 (+) \\
        Leverage             & None                              & 1998--2023 (+) \\
        Investment Intensity & 1998--2006 ($-$)                  & 1997--2023 (+) \\
        Cash Holdings        & 1999--2012 (+)                    & 1996--2023 (+) \\
        Sales (log)          & 1993--2023 (+)                    & 1993--2023 (+) \\
        Tobin Q              & 1993--1994 (+)                    & 1993 (+); 1995--2023 (+) \\
        CEO Age              & None                              & 1999--2023 (+) \\
        CEO Tenure           & None                              & 1997--2023 (+) \\
        \hline
    \end{tabularx}
    \begin{minipage}{0.98\textwidth}
    \footnotesize
    \textit{Notes:} A plus sign in parentheses denotes a positive coefficient, and a minus sign denotes a negative coefficient.
    \end{minipage}
    \label{tab:variable_significant_periods_with_pc}
\end{table}

Sales remains positive throughout 1993--2023 even after the common macro factor is included. The persistence of both the sales and PC1 coefficients indicates that the sales association remains after common macroeconomic variation is held constant, without requiring PC1 to be assigned to a single structural channel. Assets are positive in 1993--2007 but negative in 2016--2021. Within this specification, asset size is therefore consistent with an early scale-related association and a later association with mature capital, standardized operations, or greater capacity to reduce emissions intensity.

The investment result becomes more temporally concentrated. Investment intensity is negative only in 1998--2006 rather than 2000--2023, implying that part of the later baseline association is shared with common macroeconomic conditions. Cash holdings remains positive in 1999--2012, consistent with liquidity supporting activity or capacity investment on net.

In short, PC1 is positive in all or nearly all years in every paired model, consistent with a pervasive common macroeconomic component \citep{stock2002forecasting,mccracken2020fred}. The firm-level associations are not absorbed entirely by this factor. The persistent positive sales coefficient provides the clearest scale-related association conditional on PC1, while the shorter negative investment-intensity window is consistent with investment-related adjustment concentrated around the late 1990s and early 2000s. The sign reversal in assets further suggests that proxies may change their associations throughout the sample periods.

\FloatBarrier

\subsection{Time-Varying Associations in a Joint Framework}

To examine these associations jointly, the six-variable specification estimates the financial variables simultaneously: assets, leverage, investment intensity, cash holdings, sales, and Tobin's Q. Its purpose is to compare the investment and scale margins within a common facility-year sample and to measure the conditional association of each variable after the other five firm characteristics are held constant.

For the facility-level symmetric percentage change in releases, the joint model is
\begin{equation}
    \begin{aligned}
        \%\Delta &Emission_{it} = \alpha_{it} + \sum_{k\in K}\beta_{k,t}Var_{k,J(i,t)} + u_{it}, \\
        K &= \{Assets, Leverage, InvestmentIntensity, CashHoldings, \\
        & Sales, TobinQ\},
    \end{aligned}
\end{equation}
where $k\in K$ indexes the six regressors and the remaining notation follows Equation \ref{equ:baseline}.

\begin{figure}[htbp]
    \centering
    \includegraphics[width=0.95\textwidth]{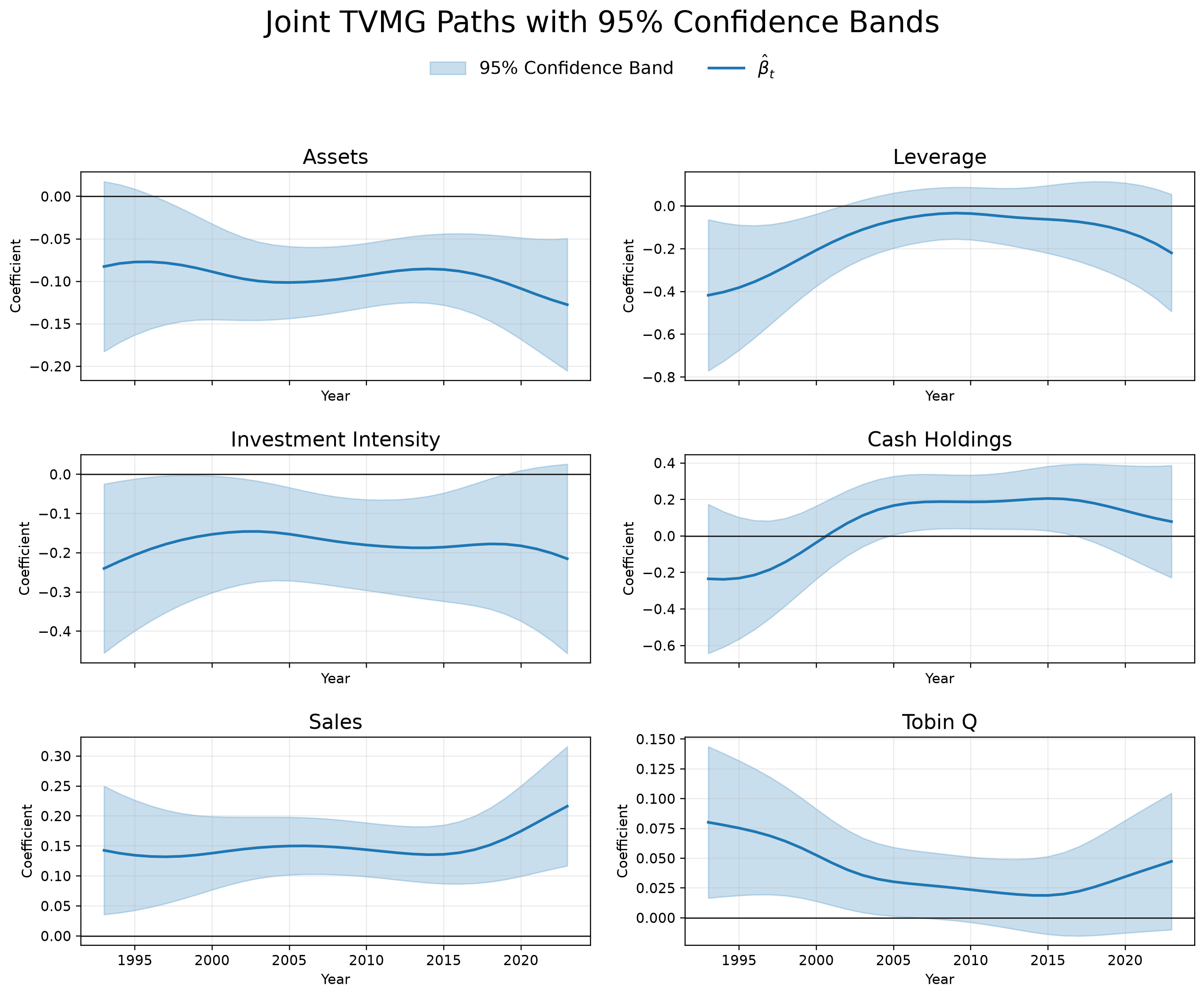}
    \caption{Joint estimation of TVMG coefficient paths for six variables together. Shaded regions are 95\% confidence bands.}
    \label{fig:multiple_tvmg_results}
\end{figure}

Requiring all six variables and the emissions-growth outcome to be observed for every estimation year leaves 3,111 facilities, 170 parent firms, and 96,441 facility-year observations over 1993--2023. Using $H=\sqrt{31}=5.57$, and 95\% confidence bands, all six coefficients have at least one significant period (Table \ref{tab:variable_significant_periods_multiple}). Their signs divide naturally between the two margins: sales, Tobin's Q, and cash holdings are positive, whereas investment intensity, assets, and leverage are negative. Figure \ref{fig:multiple_tvmg_results} displays the corresponding coefficient paths.

\begin{table}[htb]
    \centering
    \caption{Joint six-variable TVMG significant periods and directions}
    \begin{tabular}{lcc}
        \hline\hline
        \textbf{Variable} & \textbf{Direction} & \textbf{Significant Period(s)} \\
        \hline
        Assets               & Negative & 1997--2023 \\
        Leverage             & Negative & 1993--2001 \\
        Investment Intensity & Negative & 1993--2018 \\
        Cash Holdings        & Positive & 2005--2016 \\
        Sales (log)          & Positive & 1993--2023 \\
        Tobin Q              & Positive & 1993--2006 \\
        \hline
    \end{tabular}
    \label{tab:variable_significant_periods_multiple}
\end{table}

The joint model makes the scale--investment comparison sharper than the univariate results. Sales remains positive in every year, while investment intensity is negative through 2018. Conditional on the same balance-sheet and valuation controls, expansion in current operating activity is therefore associated with higher emissions growth, whereas a higher CapEx/PPE ratio is associated with lower emissions growth. The positive Tobin's Q coefficient in 1993--2006 is also consistent with a scale or growth-opportunity channel. Cash holdings becomes positive in 2005--2016, indicating that internal liquidity accompanies the expansion side of investment during that interval.

Assets turn negative from 1997--2023 once sales, investment intensity, financing structure, and valuation are held constant. This differs from the early positive univariate asset coefficient and is economically meaningful: conditional on current sales, a larger asset base need not represent higher utilization. It can instead proxy mature capital, organizational capacity, and lower emissions growth per unit of activity. The early negative leverage coefficient similarly fits a constrained-scale interpretation, in which debt burdens restrict the expansion that would otherwise raise emissions.

\paragraph{Standardized coefficients and economic magnitudes} Because the regressors use different units, we calculate $\beta_t SD(X)/SD(Y)$ using standard deviations from the common joint-estimation sample. Averaged over each variable's significant period, sales has the largest positive standardized association ($0.422$), while assets has the largest negative association ($-0.318$). The remaining average standardized coefficients are smaller: Tobin's Q ($0.084$), leverage ($-0.072$), investment intensity ($-0.033$), and cash holdings ($0.033$). Thus, after scale normalization, current sales and conditional asset size have the largest absolute standardized associations in this summary.

\begin{table}[htb]
    \centering
    \caption{Standardized coefficients and association-implied magnitudes in significant periods}
    \small
    \begin{tabularx}{\textwidth}{lccXrr}
        \hline\hline
        \textbf{Variable} & \textbf{Period} & \shortstack{\textbf{Avg. Std.}\\$\boldsymbol{\beta_t}$} &
        \textbf{Scenario} & \shortstack{\textbf{Avg. Implied}\\\textbf{Change (pp)}} &
        \shortstack{\textbf{Largest Implied}\\\textbf{Change (pp; year)}} \\
        \hline
        Assets               & 1997--2023 ($-$) & $-0.318$ & 10\% increase in assets       & $-0.92$ & $-1.21$ (2023) \\
        Leverage             & 1993--2001 ($-$) & $-0.072$ & +10 pp leverage               & $-3.09$ & $-4.17$ (1993) \\
        Investment Intensity & 1993--2018 ($-$) & $-0.033$ & +10 pp CapEx/PPE              & $-1.76$ & $-2.40$ (1993) \\
        Cash Holdings        & 2005--2016 (+)   & $ 0.033$ & +10 pp cash/assets            & $ 1.90$ & $ 2.05$ (2015) \\
        Sales                & 1993--2023 (+)   & $ 0.422$ & 10\% increase in sales        & $ 1.42$ & $ 2.06$ (2023) \\
        Tobin Q              & 1993--2006 (+)   & $ 0.084$ & +1 unit in Tobin's Q          & $ 5.46$ & $ 8.02$ (1993) \\
        \hline
    \end{tabularx}
    \begin{minipage}{0.96\textwidth}
    \footnotesize
    \textit{Notes:} A plus sign in parentheses denotes a positive coefficient, and a minus sign denotes a negative coefficient. The standardized coefficient is
    $\beta_t SD(X)/SD(Y)$. Reported changes are association-implied percentage-point changes in symmetric emissions growth, averaged within the reported significant period. ``Largest absolute'' denotes the annual association-implied change with the largest absolute value in that period, with its sign preserved. Scenario sizes differ across variables, so the magnitude columns should not be used to rank relative importance.
    \end{minipage}
    \label{tab:joint_effect_magnitudes}
\end{table}

For economic interpretation, a 10\% sales increase is associated on average with a 1.42 percentage-point increase in symmetric emissions growth during 1993--2023, whereas a 10\% increase in assets is associated with a 0.92 percentage-point decrease during 1997--2023. A 10-percentage-point increase in investment intensity is associated with a 1.76 percentage-point decrease during 1993--2018, while the same-sized increase in cash/assets is associated with a 1.90 percentage-point increase during 2005--2016. The corresponding averages are $-3.09$ percentage points for a 10-percentage-point leverage increase and $+5.46$ percentage points for a one-unit increase in Tobin's Q. At their strongest annual points within the significant periods, the association-implied changes reach $-1.21$ percentage points for assets (2023), $-4.17$ for leverage (1993), $-2.40$ for investment intensity (1993), $+2.05$ for cash holdings (2015), $+2.06$ for sales (2023), and $+8.02$ for Tobin's Q (1993).

\paragraph{Local condition numbers} For the joint model, we assess local multicollinearity by computing a scale-invariant condition number for each of the 96,441 facility-year weighted least-squares designs. Each regressor is first centered and scaled by its weighted mean and standard deviation within the local kernel window. Let $\widetilde{\mathbf{x}}_{ij,t}$ contain an intercept and these locally standardized regressors for facility $i$ in year $j$, using target year $t$. Define
\begin{equation}
    \mathbf{Q}_{it}
    =
    \sum_{j\in\mathcal{T}}
    b_{H,|j-t|}
    \widetilde{\mathbf{x}}_{ij,t}
    \widetilde{\mathbf{x}}_{ij,t}^{\prime},
    \qquad
    \kappa_{it}
    =
    \sqrt{
    \frac{\lambda_{\max}(\mathbf{Q}_{it})}
    {\lambda_{\min}(\mathbf{Q}_{it})}
    }.
    \label{equ:local_condition_number}
\end{equation}
Here $b_{H,|j-t|}$ is the Gaussian kernel weight. For the six-regressor joint specification, the median local condition number is 8.28 and the 95th percentile is 20.16. Using 30 as a diagnostic flag, only 0.60\% of the local designs exceed that value, and none exceeds 100 or is infinite. Local collinearity therefore does not appear pervasive.

Overall, the joint evidence is consistent with firm characteristics reflecting either capacity or adjustment margins. In the common sample, the positive associations for sales, valuation, and cash align more closely with a scale-expansion interpretation, whereas the negative associations for investment intensity, conditional asset size, and leverage align with investment-related adjustment or a constrained-scale interpretation.

\FloatBarrier

\subsubsection{Parsimonious Model}

\begin{figure}[htbp]
    \centering
    \begin{subfigure}{0.96\textwidth}
        \centering
        \includegraphics[width=\linewidth]{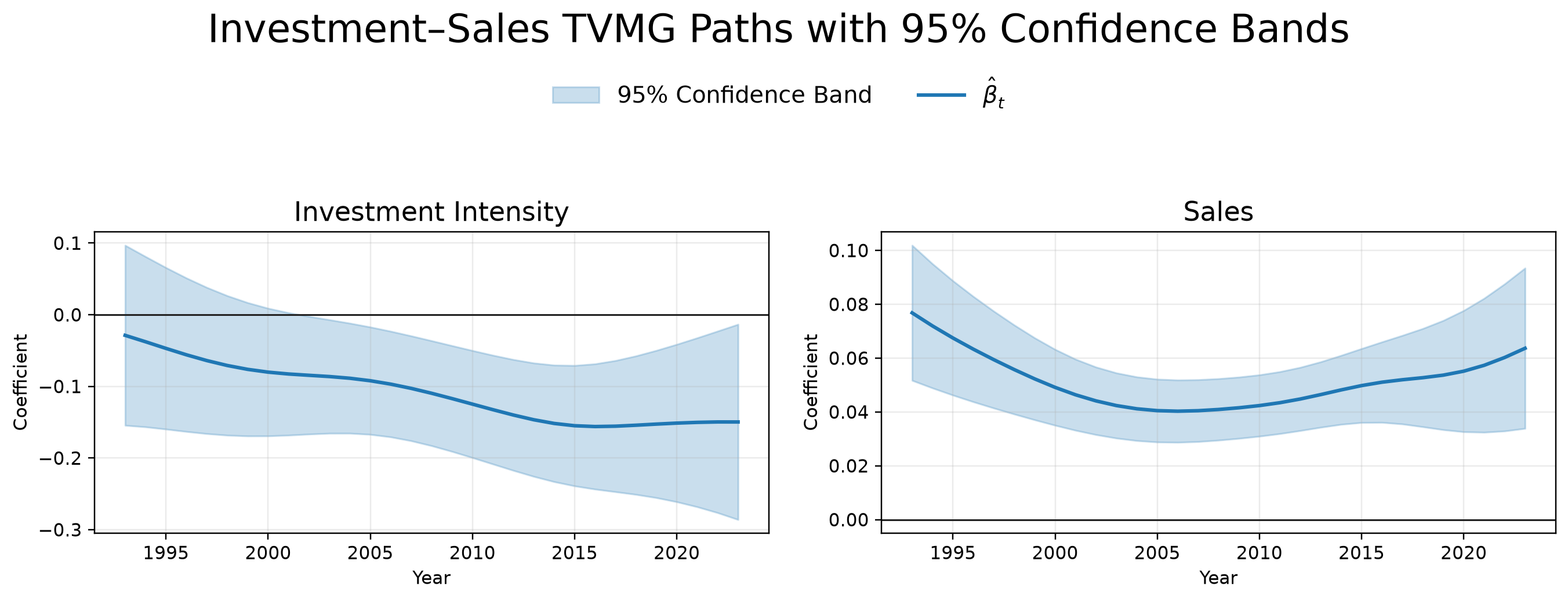}
        \caption{Facility-level model}
    \end{subfigure}

    \vspace{0.4cm}

    \begin{subfigure}{0.96\textwidth}
        \centering
        \includegraphics[width=\linewidth]{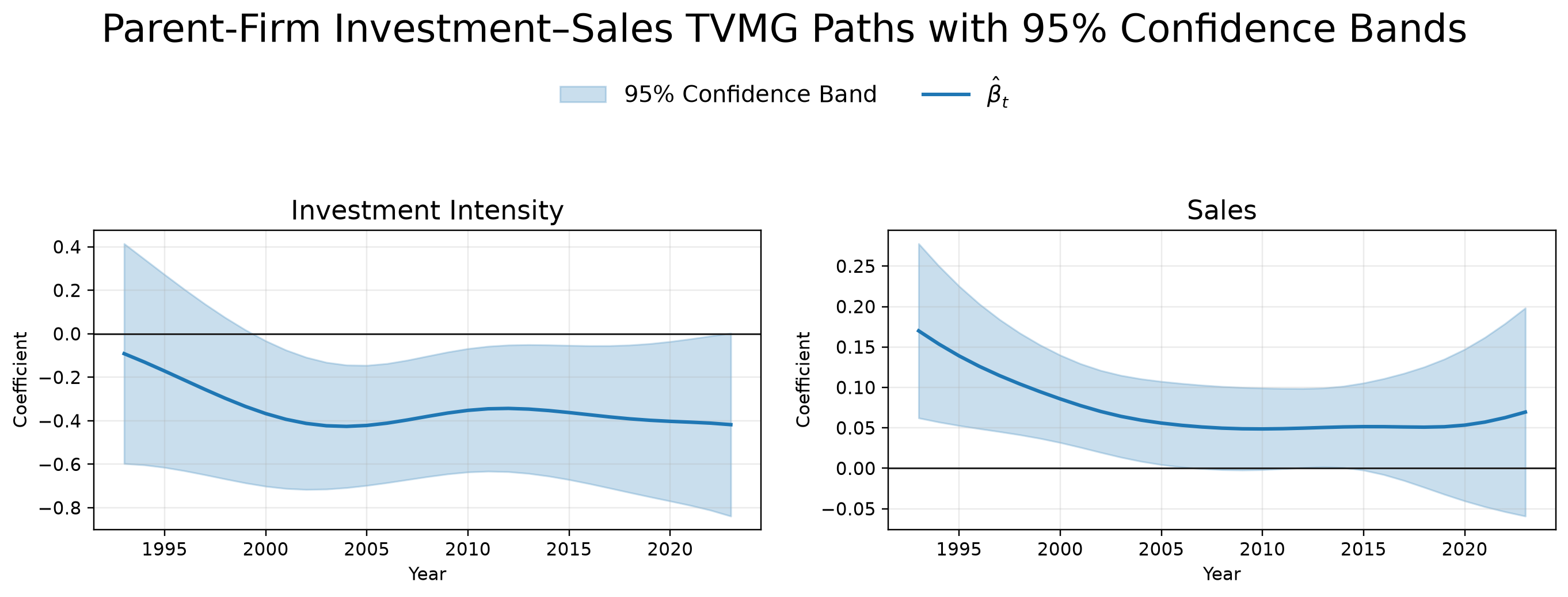}
        \caption{Parent-firm model}
    \end{subfigure}
    \caption{Investment--sales TVMG coefficient paths. Shaded regions are 95\% confidence bands.}
    \label{fig:investment_sales_joint_paths}
\end{figure}

The six-variable model conditions on a broad set of financial characteristics, but its investment coefficient may be difficult to interpret if capital expenditure and operating scale move together. We therefore first estimate a parsimonious joint model containing only investment intensity and log sales at both facility and parent-firm levels. The facility-level model is
\begin{equation}
    \%\Delta Emission_{it}
    =\alpha_{it}
    +\beta_{I,t}InvestmentIntensity_{J(i,t)}
    +\beta_{S,t}\log(Sales_{J(i,t)})+u_{it}.
    \label{equ:investment_sales_joint}
\end{equation}
Requiring the outcome and both regressors in every year produces one common sample of 4,699 facilities, 274 parent firms, and 145,669 facility-year observations. We estimate over both facility and firm level. For firm level, facility releases are first summed within each firm-year, and the bounded symmetric growth rate is then recomputed from the parent-level release totals. The balanced firm sample contains 8,494 firm-year observations. Table \ref{tab:investment_sales_joint} reports the significant periods, and Figure \ref{fig:investment_sales_joint_paths} shows the facility- and parent-firm coefficient paths.

\begin{table}[htb]
    \centering
    \caption{Investment--sales TVMG results at facility and parent-firm levels}
    \scriptsize
    \begin{tabularx}{\textwidth}{lrrXX}
        \hline\hline
        \textbf{Unit} & \textbf{Units} & \textbf{Observations} &
        \textbf{Investment Intensity} & \textbf{Sales} \\
        \hline
        Facility & 4,699 & 145,669 &
        2002--2023 ($-$) &
        1993--2023 (+) \\
        Parent firm & 274 & 8,494 &
        2000--2022 ($-$) &
        1993--2006 (+); 2012--2014 (+) \\
        \hline
    \end{tabularx}
    \begin{minipage}{0.96\textwidth}
    \footnotesize
    \textit{Notes:} A plus sign in parentheses denotes a positive coefficient, and a minus sign denotes a negative coefficient. When multiple periods are listed, each sign applies to the immediately preceding period.
    \end{minipage}
    \label{tab:investment_sales_joint}
\end{table}

At the facility level, the investment-intensity band lies below zero from 2002 through 2023, while the sales band lies above zero throughout 1993--2023. At the parent-firm level, the investment band lies below zero from 2000 through 2022, while the sales band lies above zero in 1993--2006 and 2012--2014. The negative investment association therefore survives aggregation to the level at which the financial regressors are measured, whereas the timing of the positive sales association is less persistent after firm aggregation. The two levels are not mechanically comparable because the firm specification both changes the mean-group weighting and recomputes emissions growth from aggregated releases.

Sales is a financial measure of firm scale rather than a direct measure of physical activity. On production side, we also test whether the negative investment coefficient arises because investment is concentrated in facilities or firms experiencing lower production. To separate this possibility from an investment-related technique association, we add the reporting-year TRI production/activity ratio. For facility $i$, the ratio is constructed over the 1991 core chemicals as
\begin{equation}
    PR_{it}
    =
    \frac{\sum_{c\in\mathcal{C}}W_{ict}R_{ict}}
    {\sum_{c\in\mathcal{C}}W_{ict}},
    \label{equ:production_activity_ratio}
\end{equation}
where $R_{ict}$ is a valid reported production/activity ratio between zero and three and $W_{ict}$ is positive production-related waste, converted to pounds. Chemical observations without a valid ratio or positive weight do not enter either sum.

\begin{figure}[htbp]
    \centering
    \begin{subfigure}{0.96\textwidth}
        \centering
        \includegraphics[width=\linewidth]{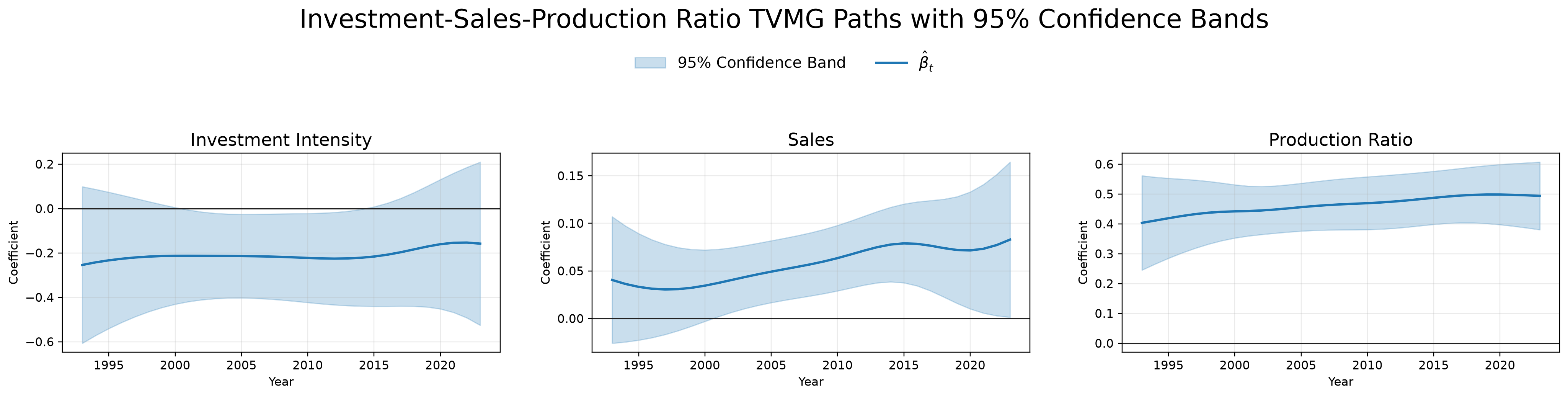}
        \caption{Facility-level model}
    \end{subfigure}

    \vspace{0.4cm}

    \begin{subfigure}{0.96\textwidth}
        \centering
        \includegraphics[width=\linewidth]{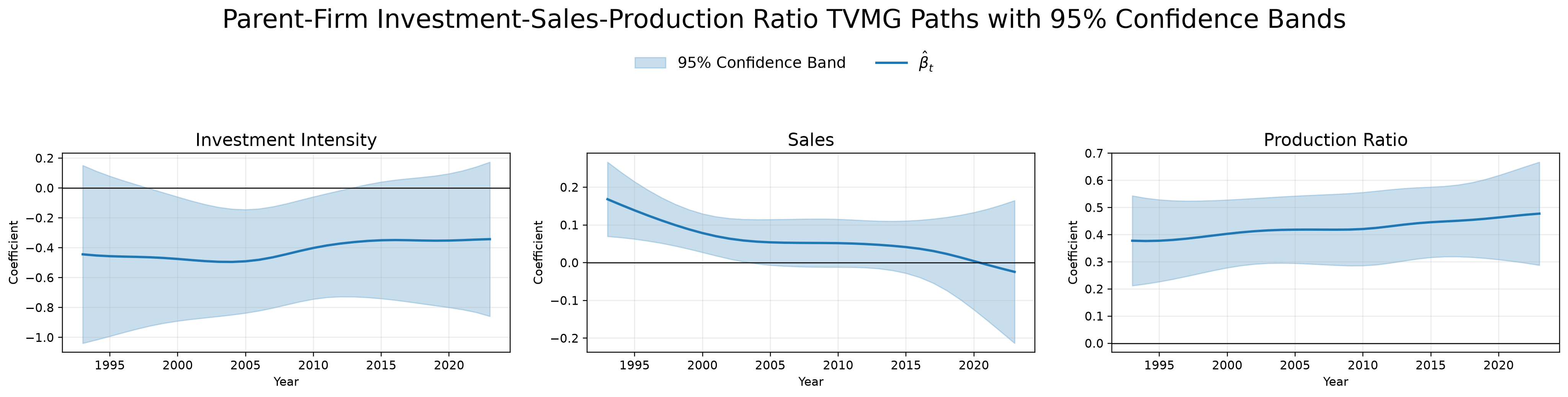}
        \caption{Parent-firm model}
    \end{subfigure}
    \caption{TVMG paths for lagged investment intensity, log sales, and the production/activity ratio. Shaded regions are 95\% confidence bands.}
    \label{fig:production_ratio_joint}
\end{figure}

At the facility level, the model includes investment intensity, sales, and the facility production/activity ratio. At the parent-firm level, facility releases are first summed within each firm-year, the bounded symmetric growth rate is recomputed from those totals, and the production-ratio numerator and denominator are aggregated across the firm's facilities. Both specifications require all three regressors and the outcome in every year. The mean-group paths and standard errors are then computed across the relevant units. Table \ref{tab:production_ratio_joint} summarizes the significant periods, and Figure \ref{fig:production_ratio_joint} shows the full paths.

\begin{table}[htbp]
    \centering
    \caption{Investment, sales, and production-ratio TVMG results}
    \scriptsize
    \begin{tabularx}{\textwidth}{lrrXXX}
        \hline\hline
        \textbf{Unit} & \textbf{Units} & \textbf{Observations} &
        \textbf{Investment Intensity} & \textbf{Sales} &
        \textbf{Production/Activity Ratio} \\
        \hline
        Facility & 601 & 18,631 &
        2001--2014 ($-$) &
        2001--2023 (+) &
        1993--2023 (+) \\
        Parent firm & 174 & 5,394 &
        1998--2012 ($-$) &
        1993--2003 (+) &
        1993--2023 (+) \\
        \hline
    \end{tabularx}
    \begin{minipage}{0.98\textwidth}
    \footnotesize
    \textit{Notes:} A plus sign in parentheses denotes a positive coefficient, and a minus sign denotes a negative coefficient.
    \end{minipage}
    \label{tab:production_ratio_joint}
\end{table}

The production/activity-ratio band lies above zero throughout 1993--2023 at both levels, providing additional evidence that higher reported production activity is associated with faster observed release growth. The investment path also retains a period below zero after both sales and production activity are held constant. At the facility level, investment is negative in 2001--2014, sales is positive in 2001--2023, and the production ratio is positive in every year. At the firm level, investment is negative in 1998--2012, sales is positive in 1993--2003, and the production ratio is again positive throughout the sample. These paths are consistent with investment capturing capital replacement, process improvement, or abatement in addition to ordinary expansion.

\FloatBarrier

\subsection{Aggregate Emissions and Macroeconomic Conditions}    \label{sec:aggregate_analysis}

The aggregate analysis asks whether common macroeconomic conditions co-move with total, air, water, and ground releases. The aggregate series sums 238,304 facility-year observations from 7,447 balanced facilities into 32 annual observations for 1992--2023. PCA is applied to 237 complete transformed FRED-QD series. Because the components are statistical factors, we do not assign them structural labels.

For this experiment, we use a time-varying model where
\begin{equation}
    \%\Delta Emission_{t} = \alpha_{t} + \sum^M_{m=1}\beta_{m,t}PC_{m,t} + u_{t}, \quad M \in \{1, 2, 3\}.
\end{equation}
We estimate one-, two-, and three-component time-varying regressions with a Gaussian kernel and $H=\sqrt{31}$. Because the aggregate series has no cross-sectional units, the mean-group aggregation in Equation \eqref{equ:mg} is not applicable. Formal inference instead uses a moving-block bootstrap with 1,000 replications, block length $\ell=\lceil T^{1/3}\rceil=4$, and 90\% pointwise percentile bands. For each replication, contiguous blocks are sampled with replacement, concatenated to length $T$, and the full local regression is re-estimated:
\begin{equation}
    \text{CI}^{\text{MBB}}_{t,\,1-\alpha}
    =
    \left[
    Q_{\alpha/2}\!(\{\hat{\beta}^{*(b)}_{t}\}_{b=1}^{B}),
    \;
    Q_{1-\alpha/2}\!(\{\hat{\beta}^{*(b)}_{t}\}_{b=1}^{B})
    \right],
\end{equation}
applied componentwise for each element of $\beta_t$. This moving-block bootstrap procedure yields confidence bands that appropriately reflect serial dependence and the local nature of TV estimation. We set $\alpha$ to 0.10 thus 90\% confidence level.

Because the confidence bands are constructed using percentile quantiles of the moving-block bootstrap distribution, they are not constrained to be centered on the point estimates. In finite samples, the sampling distribution of the kernel-based time-varying coefficient estimator can be asymmetric and affected by local smoothing bias and boundary effects, so that the empirical bootstrap distribution may be shifted relative to $\hat{\beta}_t$. As a result, the point estimate may lie asymmetrically within the confidence band or, in some cases, fall outside the band altogether. This behavior is a standard feature of percentile bootstrap confidence intervals and reflects the non-standard sampling distribution of time-varying estimators rather than an error in implementation.

\begin{figure}[htbp]
    \centering
    \includegraphics[width=0.95\textwidth]{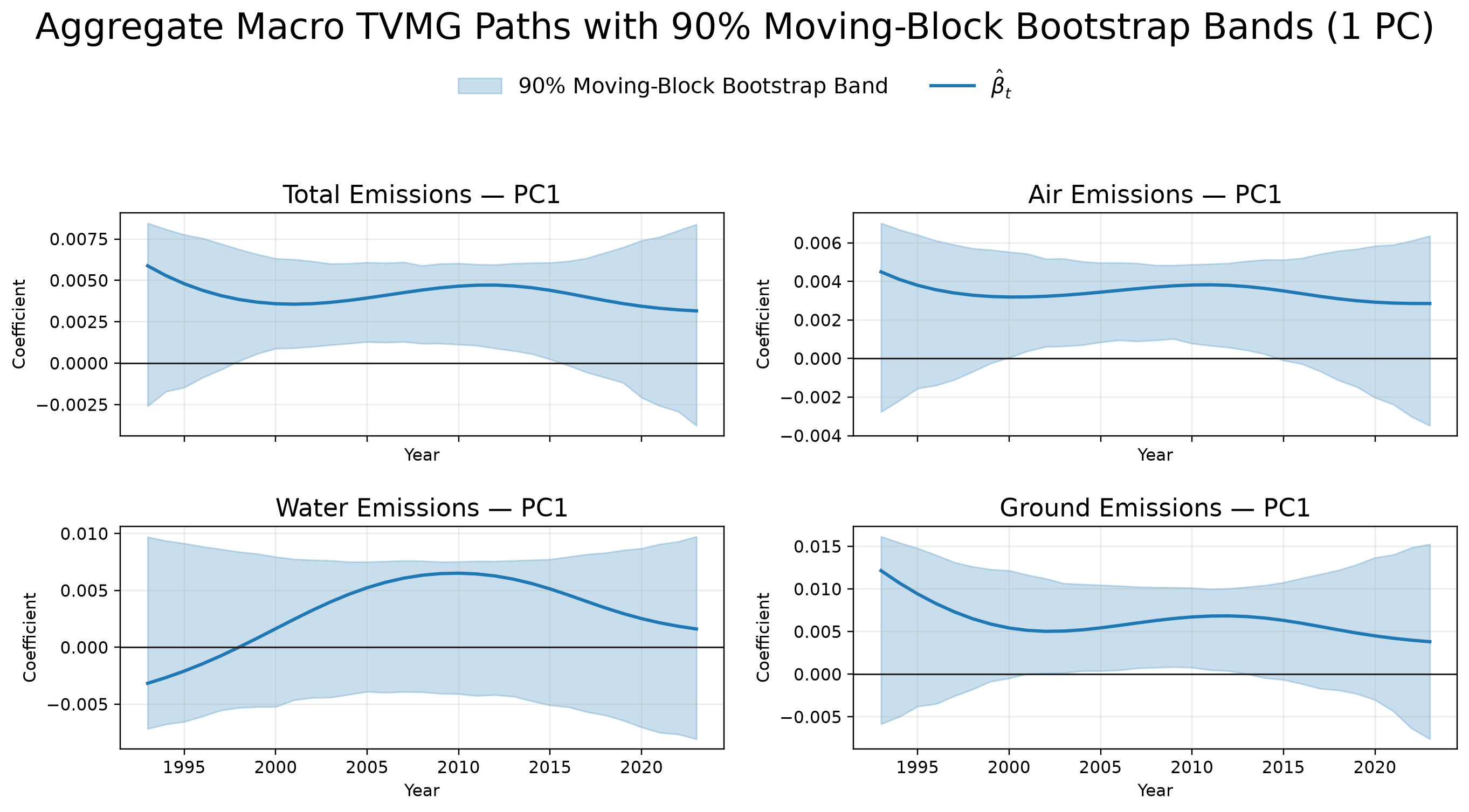}
    \caption{Time-varying coefficient estimation using aggregate emissions with the first principal component, moving-block bootstrap confidence band at 90\% level.}
    \label{fig:agg_tv_results_1pc}
\end{figure}

\begin{figure}[htbp]
    \centering
    \includegraphics[width=0.95\textwidth]{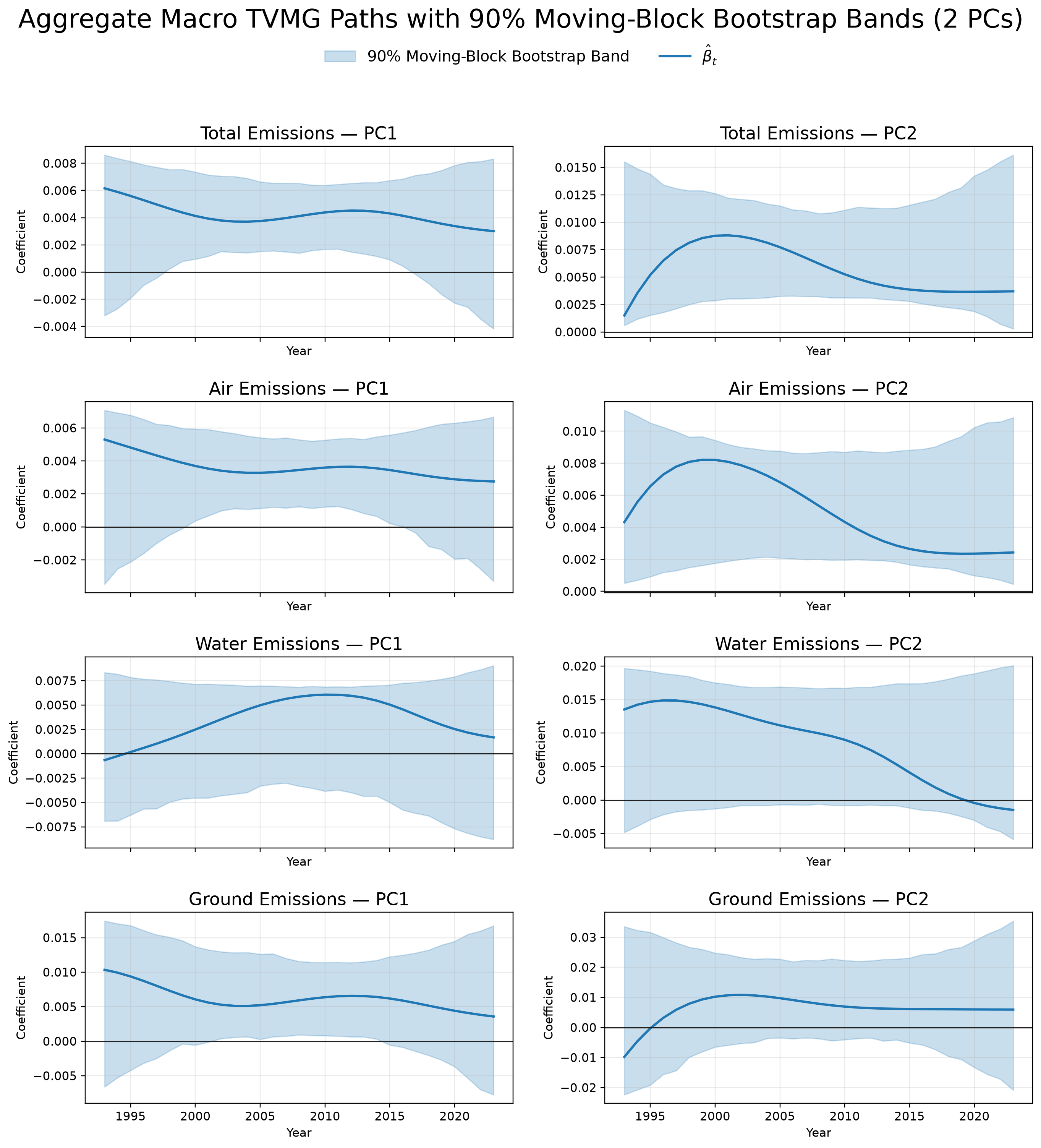}
    \caption{Time-varying coefficient estimation using aggregate emissions with the first and second principal component, moving-block bootstrap confidence band at 90\% level.}
    \label{fig:agg_tv_results_2pc}
\end{figure}

\begin{figure}[htbp]
    \centering
    \includegraphics[width=0.95\textwidth]{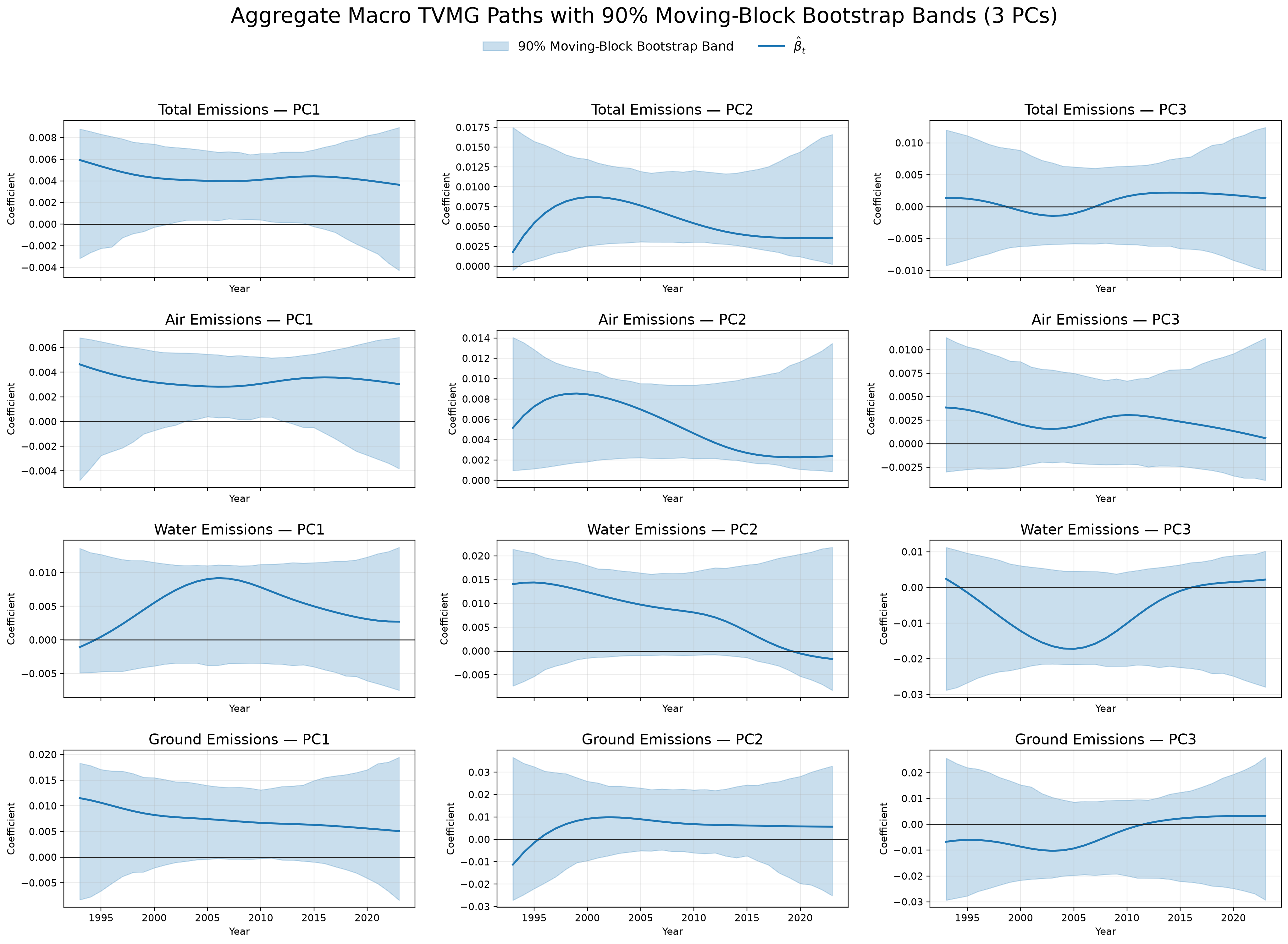}
    \caption{Time-varying coefficient estimation using aggregate emissions with the first, second, and third principal component, moving-block bootstrap confidence band at 90\% level.}
    \label{fig:agg_tv_results_3pc}
\end{figure}

Figures \ref{fig:agg_tv_results_1pc}, \ref{fig:agg_tv_results_2pc}, and \ref{fig:agg_tv_results_3pc} present the coefficient paths. Table \ref{tab:aggregate_macro_periods} reports the periods in which the 90\% moving-block bootstrap band excludes zero.

\begin{table}[htbp]
    \centering
    \caption{Aggregate emissions and FRED-QD principal components}
    \small
    \renewcommand{\arraystretch}{1.12}
    \begin{tabularx}{\textwidth}{
        @{}
        >{\raggedright\arraybackslash}p{0.14\textwidth}
        >{\centering\arraybackslash}p{0.08\textwidth}
        *{4}{>{\centering\arraybackslash}X}
        @{}
    }
        \hline\hline
        \textbf{Specification} & \textbf{PC} & \textbf{Total} &
        \textbf{Air} & \textbf{Water} & \textbf{Ground} \\
        \hline
        One PC
        & PC1 & 1998--2015 (+) & 2000--2014 (+) & None & 2001--2013 (+) \\
        \hline
        \multirow{2}{*}{Two PCs}
        & PC1 & 1998--2016 (+) & 2000--2016 (+) & None & 2002--2014 (+) \\
        & PC2 & 1993--2023 (+) & 1993--2023 (+) & None & None \\
        \hline
        \multirow{3}{*}{Three PCs}
        & PC1 & 2002--2014 (+) & 2003--2012 (+) & None & None \\
        & PC2 & 1994--2023 (+) & 1993--2023 (+) & None & None \\
        & PC3 & None & None & None & None \\
        \hline
    \end{tabularx}
    \begin{minipage}{0.96\textwidth}
    \footnotesize
    \textit{Notes:} Entries report periods in which the moving-block bootstrap 90\% pointwise confidence band excludes zero. A plus sign in parentheses denotes a positive coefficient.
    \end{minipage}
    \label{tab:aggregate_macro_periods}
\end{table}

The one-component specification shows positive mid-sample co-movement for total, air, and ground emissions, but not water emissions. When PC2 is added, it is positive throughout the sample for total and air emissions, while PC1 retains a positive mid-sample window. The three-component model preserves this total-and-air pattern with somewhat shorter PC1 windows; PC3 adds no significant period. Water emissions have no significant coefficient in any specification.

This evidence indicates common macroeconomic co-movement, especially for total and air releases. It does not identify a structural factor: PC1 and PC2 may combine aggregate conditions, and their loadings are not interpreted here. Overall, the macro results complement the facility evidence by documenting a common aggregate component, providing some evidence on aggregate scale associations.

\FloatBarrier

\subsection{Instrumental-Variable Identification Exercise}    \label{sec:iv_attempt}

The negative investment-intensity paths above are conditional associations and do not by themselves identify a causal effect. We therefore examined an excluded instrument that interacts debt-maturity exposure dated $t-2$ with the aggregate Excess Bond Premium dated $t-1$. The exercise was conducted at the unique parent-year level for first-stage diagnostics, while weak-IV-robust moments were aggregated by parent. The instrument is weak in the pooled specification, throughout the localization frontier, and under the alternative-instrument screens reported in Appendix \ref{app:iv_attempt}. The same conclusion holds in specifications that alter the sales control, add the maturity constituent, or use a predetermined pre-crisis exposure.

Weak-IV-robust Anderson--Rubin inference is correspondingly uninformative. The projected confidence set is the whole real line in every annual pointwise comparison and throughout the interior simultaneous family; the static 95\% confidence sets are also all-real in the primary and alternative-control specifications. These sets do not show that investment has no effect. They show that the available instrument does not identify a finite causal direction or magnitude. We therefore treat the IV analysis as an attempted identification exercise rather than a main result, and interpret all TVMG estimates in this paper as conditional time-varying associations. Appendix \ref{app:iv_attempt} describes the instrument, diagnostics, weak-IV-robust inference, and interpretation in detail.

\section{Conclusion}    \label{sec:conclusion}

This paper has examined the long-run evolution of the firm-level correlates of industrial toxic releases using a granular panel of U.S. manufacturing facilities spanning 1992--2023. By combining facility-level releases from the Toxics Release Inventory with parent-firm financial characteristics, managerial attributes, and aggregate macroeconomic indicators, we document how these relationships evolve across changing economic environments. Methodologically, the analysis uses a time-varying mean-group framework that allows the coefficient paths to change smoothly while accommodating persistent heterogeneity across production units.

Two broad conclusions emerge. First, the associations between firm characteristics and toxic-release growth are dynamic. Over time, the estimated coefficient paths change in magnitude, relevance, and, in some settings, direction over the sample rather than remaining constant from 1993 to 2023. Heterogeneity exists across production settings and levels of aggregation. A constant full-sample coefficient or a homogeneous response across production units would therefore conceal important features of the data.

Second, different firm characteristics have distinct empirical relationships with release growth. The clearest result is the contrast between operating scale and investment-related adjustment. Parent-firm sales is positively associated with facility emissions growth throughout the sample, whereas investment intensity is negatively associated with emissions growth over a substantial part of the period. Several measures of financial structure, valuation, liquidity, and managerial tenure are also associated with release growth during particular periods. A simple corporate finance interpretation helps organize these patterns: sales provides the clearest proxy for the operating-scale margin, while capital expenditure can reflect capacity expansion, capital replacement, process improvement, or abatement. Additional analyses provide supportive evidence for this scale--investment distinction. The opposing sales and investment paths remain in the six-variable joint specification and the parsimonious models, as well as under simultaneous inference, alternative panel constructions, and changes in aggregation and smoothing.

Taken together, these findings suggest several implications for innovation-oriented environmental strategy. They do not identify the effects of particular policies or management actions, but the variation across periods and production settings indicates that policy evaluation should not presume uniform relationships. The results motivate closer attention to firms' financial capacity, capital-replacement opportunities, and sectoral production conditions when assessing environmental adjustment. They also provide a basis for examining complementary measures that lower barriers to cleaner capital investment, despite the degree of effectiveness would require separate causal evaluation. For corporate managers, the same scale--investment distinction offers a practical framework for incorporating emissions considerations into capital budgeting and production planning. Firms may use this framework to assess strategies on expanding capacity or supporting replacement and process improvement, and to revisit that allocation as financing and production conditions change.

More broadly, the analysis illustrates the value of long-run facility evidence for studying the interaction between regulation, investment, production, and environmental performance. By documenting how firm--emissions associations evolve over several decades, the paper shows why environmental outcomes should not be treated as static compliance decisions. Future research can build on this approach by linking releases to direct measures of cleaner investment, technology adoption, and organizational change, while using stronger sources of exogenous variation to identify causal effects.

\FloatBarrier

\newpage

\bibliographystyle{agsm}
\bibliography{references}

\newpage

\begin{appendices}

\section{Economic framework}    \label{app:theory}

This appendix provides a more formal version of the economic framework used to interpret the empirical results. The objective is not to derive a complete structural model of environmental regulation. Rather, the purpose is to show how a simple investment-cost mechanism can rationalize the time-varying reduced-form relationships documented in the main text.

For expositional clarity, suppress firm subscripts and let total emissions be denoted by $\mathcal{E}_t$. We write
\begin{equation}
    \mathcal{E}_t = m_t q_t,
    \label{equ:app_cf_emissions}
\end{equation}
where $q_t$ denotes activity and $m_t$ denotes emissions intensity. Taking log differences yields
\begin{equation}
    \Delta \log \mathcal{E}_t = \Delta \log m_t + \Delta \log q_t.
    \label{equ:app_cf_growth}
\end{equation}
Equation \eqref{equ:app_cf_growth} shows that emissions growth can change either because firms expand activity or because emissions intensity changes.

We assume that firms can lower future emissions intensity through environmentally oriented investment $G_t$, so that
\begin{equation}
    m_{t+1} = (1-\rho)m_t - \chi G_t + \varepsilon_{t+1},
    \qquad 0 < \rho < 1,\quad \chi > 0.
    \label{equ:app_cf_intensity}
\end{equation}
Hence, greater environmentally oriented investment reduces future emissions intensity.

At date $t$, the firm chooses activity $q_t$ and environmentally oriented investment $G_t$ to maximize
\begin{equation}
    R(q_t;A_t) - C(q_t) - \lambda_t m_t q_t - \frac{1}{2}(\kappa + \psi_t)G_t^2 + \delta \mathbb{E}_t[V(m_{t+1})],
    \label{equ:app_cf_problem}
\end{equation}
where $A_t$ is an aggregate demand or productivity shifter, $\lambda_t$ is the effective shadow cost of emissions, and $\psi_t$ is a composite investment-cost wedge. We write
\begin{equation}
    \psi_t = \Psi(\mathcal{F}_t),
    \qquad \kappa + \psi_t > 0,
    \label{equ:app_cf_wedge}
\end{equation}
where $\mathcal{F}_t$ collects firm-level and external conditions that may affect the cost of environmentally oriented investment. These conditions can include assets, leverage, liquidity, aggregate financial conditions, and other relevant conditions. The framework does not impose a particular functional form for $\Psi(\cdot)$ or a structural sign on any individual empirical proxy.

The first-order condition for activity is
\begin{equation}
    R_q(q_t;A_t) - C_q(q_t) - \lambda_t m_t = 0,
    \label{equ:app_cf_foc_q}
\end{equation}
while the first-order condition for environmentally oriented investment is
\begin{equation}
    (\kappa + \psi_t)G_t = - \delta \chi \mathbb{E}_t[V_m(m_{t+1})].
    \label{equ:app_cf_foc_g}
\end{equation}
Since higher future emissions intensity lowers continuation value, we have $V_m(m_{t+1}) < 0$, so the right-hand side of \eqref{equ:app_cf_foc_g} is positive. Holding the expected marginal continuation value fixed, a larger investment-cost wedge reduces environmentally oriented investment.

Equation \eqref{equ:app_cf_foc_g} therefore links the cost of environmentally oriented investment to the composite wedge $\psi_t$. A larger wedge reduces $G_t$, which weakens the decline in next period's emissions intensity through \eqref{equ:app_cf_intensity}.

Because the wedge can reflect multiple firm-level and external conditions, assets, leverage, liquidity, and aggregate financial conditions may all enter $\mathcal{F}_t$. The framework does not, however, map any one of these proxies mechanically into a higher or lower $\psi_t$. Their empirical coefficients remain reduced-form conditional associations rather than estimates of separate structural investment-cost effects.

The activity choice introduces a distinct scale margin. Since sales are closely linked to output scale, variables such as sales are more likely to be positively associated with emissions growth when the scale channel dominates. At the same time, stronger environmental pressure, captured by a higher $\lambda_t$, reduces the profitability of activity for a given emissions intensity and encourages more adjustment.

The empirical interpretation of observed investment intensity further depends on its composition. Let total observed investment be denoted by $I_t$, and let the share directed toward emissions-reducing adjustment be $\omega_t \in [0,1]$. Then
\begin{equation}
    G_t = \omega_t I_t = \omega_t K_{t-1} \text{Invint}_t,
    \label{equ:app_cf_green_share}
\end{equation}
where $\text{Invint}_t = I_t/K_{t-1}$ is observed investment intensity. If $\omega_t$ rises with environmental pressure, disclosure, or monitoring, then the reduced-form coefficient on observed investment intensity becomes more negative, because a larger share of observed investment corresponds to emissions-reducing adjustment rather than ordinary capacity expansion.

Taken together, these margins make time-varying reduced-form coefficients a natural consequence of changes in the state variables $(A_t,\lambda_t,\mathcal{F}_t,\omega_t)$. Even if the firm's underlying objective function is unchanged, the reduced-form relationship between emissions growth and observed firm characteristics will vary over time as regulation matures, investment conditions change, and the composition of investment evolves.

The framework therefore rationalizes the broad empirical patterns in the main text. Sales primarily proxy the scale margin, while investment intensity is connected to the adjustment margin. Assets, leverage, cash holdings, and macroeconomic conditions may affect scale, investment costs, or both, but the framework does not assign them to a unique structural channel. The role of the empirical analysis is not to estimate the structural parameters of this framework, but to document when each of these channels is most visible in the data.

\newpage

\section{Potential overseas relocation exposure}    \label{app:relocation}

TRI records domestic on-site releases and does not show whether a firm has shifted production or emissions abroad. We therefore use Compustat foreign pretax income, $PIFO_{jt}$, as a proxy for foreign operating exposure. This variable does not measure foreign emissions or identify relocation. The exercise asks a narrower question: whether the association between a baseline covariate and domestic facility emissions differs when the parent firm has foreign income exposure.

The foreign-income file is merged to the completed facility panel before the broad coverage screen described in the main data section, using normalized GVKEY, fiscal date, and industry format. At this construction stage, the panel contains 1,271,712 facility-year rows; the subsequent coverage and parent-availability screen yields the 238,304-row broad balanced file used to define the baseline samples. The lookup is unique at the merge key, and the many-to-one merge leaves the construction-stage row count unchanged. All 247,135 facility-year rows with complete merge keys match the foreign-income file; 177,721 of those matched rows have nonmissing PIFO. Missing PIFO is assigned to the zero-exposure group under both proxy definitions, and the interaction models use the same variable-specific balanced samples as the baseline.

The main proxy is an indicator for positive foreign pretax income:
\begin{equation}
    ForeignIncomeD_{jt}
    =
    1(PIFO_{jt}>0).
\end{equation}
Missing and non-positive PIFO values set this indicator to zero. The second proxy scales non-negative foreign pretax income by lagged assets:
\begin{equation}
    ForeignIncomeIntensity_{jt}
    =
    \frac{\max(PIFO_{jt},0)}{AT_{j,t-1}}.
\end{equation}
The scaled proxy is set to zero when PIFO is missing or non-positive, or when lagged assets are missing or non-positive.

For each baseline explanatory variable $X_{J(i,t)}$, we estimate the following augmented TVMG specification:
\begin{equation}
    \%\Delta Emission_{it}
    =
    \alpha_{i,t}
    +
    \beta_{t}X_{J(i,t)}
    +
    \gamma_{t}F_{J(i,t)}
    +
    \delta_{t}\left[X_{J(i,t)} \times F_{J(i,t)}\right]
    +
    u_{it},
\end{equation}
where $F_{J(i,t)}$ is either $ForeignIncomeD_{jt}$ or $ForeignIncomeIntensity_{jt}$. The interaction coefficient $\delta_t$ measures the difference in the conditional association between $X$ and domestic emissions as foreign income exposure changes. The model uses the facility as the mean-group unit, the Gaussian kernel with $H=\sqrt{31}$, the 1993--2023 estimation window, and 95\% confidence intervals. Table \ref{tab:relocation} reports the interaction periods and variable-specific samples.

\begin{table}[htb]
\centering
\begin{threeparttable}
\caption{Foreign income exposure interaction tests}
\label{tab:relocation}
\footnotesize
\begin{tabularx}{\textwidth}{lrrXX}
\hline\hline
Variable & Facilities & Firms
& $1(PIFO>0)$ interaction
& $\max(PIFO,0)/AT_{t-1}$ interaction \\
\hline
Assets               & 5,035 & 308 & 2023 (+) & None \\
Leverage             & 5,171 & 335 & None & None \\
Investment intensity & 4,841 & 302 & None & None \\
Cash holdings        & 5,028 & 320 & None & None \\
Sales (log)          & 5,106 & 316 & None & None \\
Tobin's Q            & 3,377 & 201 & None & None \\
CEO age              & 1,939 & 80  & None & None \\
CEO tenure           & 2,923 & 119 & None & 2000--2011 (+) \\
\hline
\end{tabularx}
\begin{tablenotes}[flushleft]
\footnotesize
\item Notes: The table reports years in which the 95\% pointwise interval for $\delta_t$ excludes zero. A plus sign in parentheses denotes a positive interaction coefficient, and a minus sign denotes a negative interaction coefficient.
\end{tablenotes}
\end{threeparttable}
\end{table}

The positive-PIFO indicator produces one significant interaction: the assets interaction is positive in 2023. The corresponding baseline asset coefficient is significant in 1993--2000, so the interaction does not overlap the period that drives the baseline scale interpretation. The other seven dummy interactions are not significant in any year.

The scaled proxy also produces one significant interaction. Its interaction with CEO tenure is positive in 2000--2011, whereas the baseline tenure coefficient is significant in 2018--2023. The other seven scaled interactions are not significant.

Neither proxy moderates sales or investment intensity in any year. Foreign income exposure therefore does not account for the positive sales and negative investment-intensity pattern in the baseline. The same absence of interaction applies to leverage, cash holdings, Tobin's Q, and CEO age. Overall, foreign income exposure do not systematically separate the baseline domestic emissions associations.

\newpage

\section{Sectoral analysis}    \label{app:sector}

The main sectoral classification uses the primary NAICS recorded for each TRI facility. This choice matches the industry definition to the establishment that generates the releases. It also avoids assigning every facility of a diversified firm to the firm's principal line of business. A facility enters the exercise only if it has the same parent and the same nonmissing NAICS3 in every year from 1992 through 2023. Candidate firms must own at least three persistent facilities, and a facility industry must contain at least ten candidate firms in every year before variable-specific balancing.

The ownership screen identifies 7,369 persistent firm-facility pairs and 487 candidate firms. Of these persistent facilities, 6,799 have a valid and unchanged NAICS3 over the full ownership window. Seventeen industries pass the ten-firm threshold. They contain 6,020 facilities owned by 480 distinct firms and contribute 186,620 facility-year observations before variable-specific completeness restrictions. Table \ref{tab:sector_facility_coverage} reports the retained industry coverage. Firm counts are not additive because 311 firms own facilities in more than one retained facility industry. Each facility, however, belongs to only one NAICS3.

\begin{table}[htb]
\centering
\caption{Coverage of retained facility NAICS industries}
\label{tab:sector_facility_coverage}
\small
\begin{tabularx}{\textwidth}{cXrr}
\hline\hline
NAICS3 & Industry & Candidate firms & Persistent facilities \\
\hline
311 & Food manufacturing & 42 & 461 \\
313 & Textile mills & 19 & 46 \\
321 & Wood product manufacturing & 21 & 158 \\
322 & Paper manufacturing & 36 & 160 \\
323 & Printing and related support activities & 22 & 98 \\
324 & Petroleum and coal products manufacturing & 29 & 142 \\
325 & Chemical manufacturing & 160 & 1,131 \\
326 & Plastics and rubber products manufacturing & 84 & 277 \\
327 & Nonmetallic mineral product manufacturing & 42 & 230 \\
331 & Primary metal manufacturing & 82 & 359 \\
332 & Fabricated metal product manufacturing & 171 & 928 \\
333 & Machinery manufacturing & 110 & 536 \\
334 & Computer and electronic product manufacturing & 82 & 307 \\
335 & Electrical equipment and component manufacturing & 54 & 282 \\
336 & Transportation equipment manufacturing & 111 & 644 \\
337 & Furniture and related product manufacturing & 18 & 110 \\
339 & Miscellaneous manufacturing & 62 & 151 \\
\hline
Unique retained sample & & 480 & 6,020 \\
\hline
\end{tabularx}
\begin{minipage}{0.94\textwidth}
\footnotesize
\textit{Notes:} Candidate firms own at least three facilities that remain with the same parent throughout 1992--2023. A retained facility also has a valid and unchanged three-digit primary NAICS throughout that window. Each industry contains at least ten candidate firms in every year. Firm counts across rows are not additive because a parent may own facilities in several industries.
\end{minipage}
\end{table}

All eight univariate specifications are estimated separately within each retained industry, using the fixed Gaussian bandwidth $H=\sqrt{31}$ and the same 95\% confidence intervals as in the baseline. This produces 136 industry-variable specifications. Seventy-four have at least one significant period, 62 have none, and no specification is lost because of an empty variable-specific balanced sample. Table \ref{tab:sector_scale_investment} reports the periods for sales and investment intensity, the two variables used to organize the main interpretation.

\begin{landscape}
\begin{table}[htbp]
\centering
\caption{Sales and investment-intensity associations by facility industry}
\label{tab:sector_scale_investment}
\scriptsize
\begin{tabularx}{\linewidth}{clcXcX}
\hline\hline
NAICS3 & Industry & \shortstack{Sales\\Direction} & \shortstack{Sales\\Period(s)} &
\shortstack{Investment\\Direction} & \shortstack{Investment\\Period(s)} \\
\hline
311 & Food manufacturing & Positive & 2008--2020 & --- & None \\
313 & Textile mills & Positive & 2003--2011 & --- & None \\
321 & Wood product manufacturing & Positive & 2000--2011 & Negative & 2009--2017 \\
322 & Paper manufacturing & --- & None & --- & None \\
323 & Printing and related support activities & Positive & 1993; 2016--2017 & Positive & 1993--1999; 2012--2016 \\
324 & Petroleum and coal products manufacturing & --- & None & --- & None \\
325 & Chemical manufacturing & Positive & 1993--2016 & --- & None \\
326 & Plastics and rubber products manufacturing & --- & None & --- & None \\
327 & Nonmetallic mineral product manufacturing & Positive & 1993--1996; 2009--2013 & Negative & 2012--2023 \\
331 & Primary metal manufacturing & --- & None & --- & None \\
332 & Fabricated metal product manufacturing & Positive & 1993--1998; 2011--2016 & --- & None \\
333 & Machinery manufacturing & Positive & 1994--2023 & Negative & 2000--2017 \\
334 & Computer and electronic product manufacturing & Positive & 1993--2013 & --- & None \\
335 & Electrical equipment and component manufacturing & Positive & 1999--2015 & --- & None \\
336 & Transportation equipment manufacturing & Positive & 1993--2004; 2013--2023 & --- & None \\
337 & Furniture and related product manufacturing & Negative; Positive & 1995--2003; 2017--2023 & --- & None \\
339 & Miscellaneous manufacturing & Positive & 1993--2006 & Negative & 2003--2007 \\
\hline
\end{tabularx}
\begin{minipage}{0.95\linewidth}
\footnotesize
\textit{Notes:} A single direction applies to every listed period unless multiple directions are shown; in that case, directions and periods correspond in the order listed.
\end{minipage}
\end{table}
\end{landscape}

Sales has a positive significant period in 13 industries. Furniture manufacturing is the only industry in which it also has a negative period, from 1995 through 2003, before becoming positive in 2017--2023. The broad prevalence of positive sales coefficients is consistent with the activity margin: greater parent-firm sales tend to accompany faster emissions growth at the firm's facilities. The result is not universal, since four industries have no significant sales period and the timing differs across those with a positive association.

Investment intensity is significant in five industries. Its coefficient is negative in wood products, nonmetallic minerals, machinery, and miscellaneous manufacturing. Each of these industries also has a positive sales period. This pattern separates two reduced-form margins within the same production setting. Sales records the association between activity and emissions, while CapEx relative to PPE can capture capital replacement, process improvement, or other investment associated with lower subsequent emissions growth. For an interpretation example, the contrast is clearest in machinery, where sales is positive over 1994--2023 and investment intensity is negative over 2000--2017. Printing is different: investment intensity is positive in 1993--1999 and 2012--2016, consistent with capital spending being associated with expansion rather than lower emissions intensity in that sample.

The sectoral results are best read as evidence that heterogeneity exists along production lines and is consistent with sector-specific balances between scale expansion and investment-related adjustment.

\newpage

\section{Instrumental-variable identification exercise}    \label{app:iv_attempt}

This appendix examines whether the negative conditional association between investment intensity and toxic-release growth admits a causal interpretation. The exercise does not yield usable identification: the proposed instrument is weak in pooled, local, and alternative specifications, and weak-IV-robust confidence sets do not identify a finite sign or magnitude.

\subsection{Design and estimand}

For an outcome dated $t$, $y_{it}$ is the consecutive symmetric change in total TRI releases from $t-1$ to $t$, and $x_{it}$ is investment intensity measured at $t-1$. The fixed-parent time-varying sample contains 210,072 facility-year observations for 7,842 facilities and 778 parents over 1993--2023. Facilities must remain with one parent throughout the model window and have at least 12 complete observations. The sample is not screened by estimated first-stage statistics.

Pooled and static diagnostics use 18,137 parent-year observations from 780 parents, obtained by averaging facility outcomes within parent-year. The time-varying AR procedure instead sums facility moments within the fixed parent before estimating the parent-level covariance. Both constructions avoid treating repeated firm-level variables as independent observations.

The partial time-varying specification is
\begin{equation}
    y_{it}
    =
    \alpha_i+x_{it}\beta_i(s_t)+\mathbf{w}_{it}'\boldsymbol{\gamma}_i+u_{it},
    \qquad s_t=t/T,
    \label{equ:iv_partial_tv}
\end{equation}
where $x_{it}$ is lagged investment intensity and $\mathbf{w}_{it}$ contains contemporaneous log sales and leave-one-parent-out cross-sectional averages of investment and the parent-level outcome. In the leading point specification, the intercept and nuisance coefficients are estimated over the full span, while only the investment slope and its local derivative vary with the evaluation date. Contemporaneous sales conditions on current production scale but may also be a post-investment mediator, so the coefficient is not a total investment effect even if the instrument were valid.

We consider three point specifications. The median coefficient path and condition number are 0.996 and $8.25\times10^{7}$ for a fixed first stage with a fully time-varying second stage, $-1.039$ and 1.84 for a fixed first stage with a partial time-varying second stage, and $-0.532$ and 13.63 when both stages vary locally. We use the fixed-first-stage, partial-TV model as the leading point specification because it is better conditioned. Conditioning does not resolve weak identification. Moreover, the orthogonality condition needed to equate the IV-strength-weighted target with the equal-parent mean path is not verified. We therefore interpret the target as an IV-strength-weighted average path.

The proposed excluded instrument interacts firm-specific debt-maturity exposure with an aggregate credit shock:
\begin{equation}
    z_{g,t}
    =
    \operatorname{Win}_{1,99}\!\left(
    \frac{DD1}{DLTT}
    \right)_{g,t-2}
    \times
    \operatorname{Win}_{1,99}\!\left(EBP_{t-1}\right),
    \label{equ:iv_instrument}
\end{equation}
Here $DD1/DLTT$ is short-term debt-maturity exposure and $EBP$ is the annual mean Excess Bond Premium. Missing $DD1$ or missing or non-positive $DLTT$ leaves exposure undefined. Calendar matching uses maturity at $t-2$, EBP at $t-1$, and investment at $t-1$; reconstruction of the interaction has zero numerical error. The primary static model includes parent and year effects, the maturity main effect, contemporaneous sales, and leave-one-parent-out CCE controls. Year effects absorb the EBP main effect. Predetermined timing does not establish exclusion, since credit conditions may affect emissions through production, maintenance, closure, input choice, or pollution-control spending outside measured CapEx/PPE.

\subsection{First-stage evidence and alternative designs}

Table \ref{tab:iv_strength} summarizes the pooled and local first stages. The pooled effective F is 2.46, and the interaction F falls to 2.07 when the maturity constituent enters separately. The median full-span parent-specific Montiel Olea--Pflueger effective F is 1.26, with a 90th percentile of 9.31. Across dates, the median parent-specific local effective F ranges from 2.03 to 2.57, and 82.9\% to 90.6\% of finite parent-date statistics are below 10. The bandwidth frontier remains weak: its highest median effective F is 3.27 at $H=12$. No first-stage statistic is used to select facilities or parents.

\begin{table}[htbp]
    \centering
    \caption{Pooled and local first-stage diagnostics}
    \small
    \begin{tabularx}{\textwidth}{lXrr}
        \hline\hline
        \textbf{Diagnostic} & \textbf{Specification} & \textbf{Effective F} & \textbf{Partial $R^2$} \\
        \hline
        Pooled & Frozen parent-year sample & 2.46 & $9.07\times10^{-5}$ \\
        Primary interaction & Pooled with maturity constituent & 2.07 & $9.83\times10^{-5}$ \\
        Local constant & $H=\sqrt{31}$; median across dates & 1.65 & $1.23\times10^{-4}$ \\
        Local linear & $H=\sqrt{31}$; median across dates & 0.49 & $2.31\times10^{-4}$ \\
        Best bandwidth-frontier median & Local constant, $H=12$ & 3.27 & $1.35\times10^{-4}$ \\
        \hline
    \end{tabularx}
    \begin{minipage}{0.96\textwidth}
    \footnotesize
    \textit{Notes:} The pooled statistics use parent-cluster inference. Local rows keep parent and year effects and nuisance coefficients pooled while localizing the excluded coefficient. Local entries report medians across dates and are not directly interchangeable with the distribution of parent-specific MOP effective-F statistics.
    \end{minipage}
    \label{tab:iv_strength}
\end{table}

The three alternative instruments are also weak. Table \ref{tab:iv_alternatives} reports the largest effective F across candidate-specific and common samples, with and without 1st/99th-percentile clipping.

\begin{table}[htbp]
    \centering
    \caption{Alternative-instrument relevance screens}
    \small
    \begin{tabularx}{\textwidth}{XrX}
        \hline\hline
        \textbf{Candidate} & \textbf{Maximum effective F} & \textbf{Primary exclusion concern} \\
        \hline
        Leverage at $t-1$ & 3.41 & Can affect production, financial constraints, and pollution directly rather than only through investment. \\
        Fixed 1993 machinery exposure $\times$ bonus rate at $t-1$ & 0.41 & Captures a general tax-sensitive investment incentive, not environmental-purpose investment. \\
        $PPEGT-PPENT$ at $t-1$ & 0.95 & Reflects capital stock, depreciation, and vintage composition with direct scale and technique pathways. \\
        \hline
    \end{tabularx}
    \begin{minipage}{0.96\textwidth}
    \footnotesize
    \textit{Notes:} The maximum of 3.41 is the clipped leverage candidate on the strict common sample. These statistics rank relevance only; none establishes an exclusion restriction.
    \end{minipage}
    \label{tab:iv_alternatives}
\end{table}

The pre-crisis design uses 2004 debt data to define exposure before the 2007--2009 crisis. The resulting sample has 6,254 parent-years and 635 parents; the parent-cluster and two-way F statistics are 1.32 and 1.47, and both 95\% AR sets are all-real. Lead tests do not reject pretrends, but this non-rejection does not establish exclusion and the pseudo-crisis windows remain weak. Replacing point exposure with its 2004--2006 average also produces unbounded or all-real AR sets.

\subsection{Weak-IV-robust inference}

For target year $r$, projected AR inference uses the original endogenous regressor and excluded instrument. The intercept and nuisance coefficients are estimated over the full span, the local derivative $b_1$ is profiled for each candidate $b_0$, and facility moments are aggregated within parent. The pointwise confidence set is
\begin{equation}
    \mathcal{C}_{r}(1-\alpha)
    =
    \left\{
    b_0:
    \inf_{b_1} AR_{MG,r}(b_0,b_1)
    \leq c_{1-\alpha}
    \right\}.
    \label{equ:iv_projected_ar}
\end{equation}
The covariance matrix is recomputed for each candidate, and the search grid expands until tail behavior is classified. Simultaneous inference over 1995--2021 uses 999 common parent resamples and re-profiles $b_1$ within each draw. The bootstrap critical value is 7.38; the reported value is the larger Bonferroni safeguard, 11.20. A circular three-year time-block critical value of 6.45 is retained as a sensitivity. The AR-moment dependence diagnostic flags 19 dates, but none of these adjustments yields a finite confidence region. Table \ref{tab:iv_ar_summary} reports the resulting pointwise, simultaneous, and static sets.

\begin{table}[htbp]
    \centering
    \caption{Weak-IV-robust Anderson--Rubin results}
    \small
    \begin{tabularx}{\textwidth}{lclX}
        \hline\hline
        \textbf{Design} & \textbf{Confidence} & \textbf{Coverage in time} & \textbf{Result} \\
        \hline
        Projected pointwise path & 90\% & 1993--2023, 31 dates & 31 of 31 sets are the whole real line; zero is never rejected. \\
        Simultaneous projected family & 90\% & 1995--2021, 27 dates & 27 of 27 sets are the whole real line; zero is never rejected. \\
        Static primary, parent covariance & 95\% & Full sample & $[-\infty,+\infty]$. \\
        Static primary, parent-plus-year score & 95\% & Full sample & $[-\infty,+\infty]$. \\
        Static sales-control variants & 95\% & Strict common sample & All-real with lagged sales or no sales. \\
        Predetermined pre-crisis design & 95\% & 2003--2012 & All-real or unbounded across exposure definitions and covariance choices. \\
        \hline
    \end{tabularx}
    \label{tab:iv_ar_summary}
\end{table}

The primary static 90\% parent-aggregated set is already unbounded, $[-\infty,0.093]\cup[0.746,+\infty]$, and expands to the whole real line at 95\%. The standardized reduced-form 95\% interval is $[-8.94\times10^{-5},3.23\times10^{-4}]$, but it concerns financing stress rather than the investment channel. An all-real or unbounded AR set indicates that the data do not identify a finite causal sign or magnitude; it is not evidence of a zero effect. The point estimates and reduced-form null results therefore remain non-causal diagnostics.

\subsection{Methodological extensions and Monte Carlo calibration}

The empirical implementation evaluates several methods that can improve prediction or numerical stability without creating excluded variation. Cross-fitted controls alone have out-of-fold $R^2=0.0536$, while adding the original economic instrument raises it only to 0.0544, an increment of 0.00084. Ridge models with many predetermined financial predictors attain out-of-fold $R^2$ values of 0.276 and 0.295, but fitted investment based on those predictors is not automatically a valid instrument. A nuisance-path penalty is also examined; generalized cross-validation selects a penalty of zero. Direct-effect sensitivity regions are reported for explicit bounds, but they omit sampling uncertainty and do not supersede projected AR inference.

To make the calibration explicit, the core experiment formalizes the short Monte Carlo used to motivate the weak-IV procedure. It sets $T=31$ and uses 50 parent firms. Except in the clustered design, each parent contributes one unit. The true mean path is linear,
\begin{equation}
    \beta_0(s)=0.50+0.60(s-0.50),
    \qquad s=t/T,
    \label{equ:iv_mc_linear_path}
\end{equation}
so the local-linear approximation is exact. The excluded instrument follows $z_{it}=0.4z_{i,t-1}+e_{it}$, the endogenous regressor is $x_{it}=\pi_i z_{it}+v_{it}$, and the structural disturbance has correlation 0.6 with $v_{it}$. Unit coefficients contain an additive random deviation with standard deviation 0.15. Estimation uses a Gaussian kernel with $H=\sqrt{T}$ and nominal 90\% inference.

Four designs isolate different problems. In the strong independent design, $\pi_i=1$. In the weak independent design, $\pi_i=0.18$. The strong parent-clustered design gives each parent two facilities, sets $\pi_i=1$, and adds the same parent-year shock to the two facilities, with loadings 0.7 in the structural disturbance and 0.4 in the first-stage disturbance. Finally, the strength-correlated-heterogeneity design draws $\pi_i$ uniformly from $[0.55,1.45]$ and adds $0.7(\pi_i-\bar\pi)$ to the unit-specific coefficient deviation. The last design is a deliberate estimand stress test because it violates the orthogonality between coefficient heterogeneity and first-stage strength required for the simple mean path to be the IV target.

The experiment compares two point estimators. The fixed-first-stage estimator fits $x_{it}$ on $(1,z_{it})$ over the full unit time series and then applies a kernel-weighted partial local-linear second stage to fitted investment. The fully local estimator uses $(1,x_{it},x_{it}v_{tr})$ as regressors and $(1,z_{it},z_{it}v_{tr})$ as instruments within each local window. Mean-group estimates are equally weighted across parents. Conventional cross-parent Wald intervals are computed for both estimators. The projected AR statistic instead removes unit intercepts by weighted demeaning, sums facility moments within parent, estimates the covariance across parents, and minimizes the two-moment quadratic form over the local derivative before comparison with the 90\% $\chi^2_2$ critical value. Results are evaluated at the first, middle, and last dates. AR power is rejection of a false coefficient 0.40 above the truth in the strong designs and 0.50 above the truth in the weak design.

Each design uses 1,000 replications, rather than the 100-replication preliminary exercise. Thus, the Monte Carlo standard error of a coverage estimate near 0.90 is approximately 0.0095. Table \ref{tab:iv_mc_core} reports the full higher-replication results.

\begin{landscape}
\begin{table}[htbp]
\centering
\caption{Core Monte Carlo results: point estimation and projected AR inference}
\label{tab:iv_mc_core}
\scriptsize
\begin{tabularx}{\linewidth}{Xlrrrrrrrr}
\hline\hline
Design & Date & Global $F$ & Local $F$ & RMSE fixed & RMSE local & Wald fixed & Wald local & AR coverage & AR power \\
\hline
Strong, independent
    & First  & 32.698 & 15.878 & 0.138 & 66.930 & 0.884 & 0.919 & 0.967 & 0.996 \\
    & Middle & 32.698 & 17.131 & 0.051 &  2.561 & 0.886 & 0.703 & 0.972 & 1.000 \\
    & Last   & 32.698 & 18.674 & 0.151 &  8.838 & 0.885 & 0.898 & 0.958 & 0.998 \\
\hline
Weak, independent
    & First  & 1.162 & 2.092 & 18.498 & 42.828 & 0.925 & 0.876 & 0.966 & 0.161 \\
    & Middle & 1.162 & 1.220 & 12.086 & 18.574 & 0.912 & 0.858 & 0.959 & 0.914 \\
    & Last   & 1.162 & 2.307 & 26.599 & 49.710 & 0.928 & 0.856 & 0.959 & 0.221 \\
\hline
Strong, parent clusters
    & First  & 28.183 & 13.976 & 0.125 & 37.498 & 0.869 & 0.906 & 0.957 & 0.998 \\
    & Middle & 28.183 & 14.892 & 0.055 &  3.364 & 0.761 & 0.373 & 0.955 & 1.000 \\
    & Last   & 28.183 & 16.260 & 0.130 & 19.927 & 0.873 & 0.895 & 0.958 & 1.000 \\
\hline
Strength-correlated heterogeneity
    & First  & 32.012 & 15.497 & 0.238 & 107.494 & 0.901 & 0.936 & 0.966 & 0.963 \\
    & Middle & 32.012 & 16.640 & 0.424 &   1.323 & 0.918 & 0.820 & 0.933 & 1.000 \\
    & Last   & 32.012 & 18.290 & 0.709 &   7.819 & 0.887 & 0.930 & 0.950 & 0.993 \\
\hline
\end{tabularx}
\begin{minipage}{0.97\linewidth}
\footnotesize
\textit{Notes:} Each design uses 1,000 replications. ``Fixed'' denotes a full-sample first stage followed by a partial local-linear second stage; ``local'' denotes local first and second stages. Wald and AR columns report empirical coverage of nominal 90\% procedures. Global and local $F$ statistics are medians. AR power is rejection of the false coefficient described in the text.
\end{minipage}
\end{table}
\end{landscape}

Table \ref{tab:iv_mc_core} shows three patterns. First, in the strong independent design the fixed-first-stage estimator is much more stable than the fully local IV ratio. Its RMSE is between 0.051 and 0.151, whereas the local estimator's RMSE rises to 66.930 at the first date and 8.838 at the last date. A median local $F$ above 15 therefore does not eliminate heavy-tailed unit-level IV ratios or boundary instability. The local Wald interval also covers only 0.703 at the midpoint, while projected AR coverage remains between 0.958 and 0.972.

Second, for IV analysis, parent dependence must be respected. With a strong instrument but two facilities sharing a parent shock, conventional Wald coverage at the midpoint falls to 0.761 for the fixed estimator and 0.373 for the local estimator. The parent-aggregated projected AR procedure maintains coverage between 0.955 and 0.958.

Third, weak identification destroys the usefulness of both point estimators. In the weak design, the global median $F$ is 1.162, the local median $F$ ranges from 1.220 to 2.307, and fixed-estimator RMSE ranges from 12.086 to 26.599. Projected AR coverage remains close to nominal, between 0.959 and 0.966, but power against the false value falls to 0.161 at the first date and 0.221 at the last date. The higher midpoint power of 0.914 does not alter the boundary result: weak-IV-robust inference can preserve size while providing little information about direction or magnitude in parts of the path.

The strength-correlated-heterogeneity design is interpreted separately. In the 1,000-replication run, projected AR coverage ranges from 0.933 to 0.966, so this particular calibration does not display a large finite-sample size failure. That outcome does not validate the orthogonality condition or show that the AR statistic recovers the unweighted mean path when heterogeneity is related to first-stage strength. It remains an estimand problem: AR methods address weak-inference distortions conditional on the maintained moment restriction; they do not decide whether the aggregate IV moment targets the simple mean, a strength-weighted mean, or some other average.

The core design is deliberately simple. The extended experiment adds the features it omits and again uses $T=31$, 50 parents, 1,000 replications, and two excluded instruments, each following an AR(1) process with coefficient 0.4. The true path is nonlinear,
\begin{equation}
    \beta_0(s)=0.50+0.25\sin\{2\pi(s-0.50)\}.
    \label{equ:iv_mc_nonlinear_path}
\end{equation}
The observed common factor follows an AR(1) process with coefficient 0.5 and enters the first and second stages with loadings 0.5 and 0.7. Two additional latent persistent common shocks enter the endogenous regressor and structural disturbance through heterogeneous loadings drawn from $[0.35,0.75]$. The first latent shock has autoregressive coefficient 0.65; the second has coefficient 0.55 and also loads 0.35 on the first. Instrument coefficients are heterogeneous, with $\pi_{1i}\sim U[0.65,1.15]$ and $\pi_{2i}\sim U[0.25,0.75]$.

The nuisance specification includes the observed factor and leave-one-parent-out cross-sectional means of $x$ and $y$, so a parent's own outcome and endogenous regressor do not enter its CCE controls. A parent-demeaned pooled first stage uses both instruments, the observed factor, and the leave-one-out CCE controls to generate fitted investment. For point estimation, the bandwidth is selected from $\{0.75\sqrt{T},\sqrt{T},1.5\sqrt{T}\}$; the median selected value is 8.352. Projected AR inference deliberately undersmooths with the fixed bandwidth $0.5\sqrt{T}=2.784$. Simultaneous coverage is assessed over seven prespecified dates, $t\in\{4,8,12,16,20,24,28\}$, using 199 bootstrap draws in every replication. The leading method combines leave-one-out CCE, parent resampling, derivative profiling, and a finite-family Bonferroni critical value as a lower bound on the bootstrap supremum critical value.

The generated-pooled fixed estimator has RMSE 0.116 and the fully local multiple-IV estimator has RMSE 0.103 in this stronger, nonlinear design. The mean fitted first-stage coefficients are 0.903 and 0.500. Pointwise AR coverage for the leading procedure averages 0.911 across the seven evaluation dates: it is 0.939, 0.896, 0.919, 0.931, 0.895, 0.884, and 0.914 from the first through the seventh evaluation date. Average coverage is 0.927 at the two outer dates and 0.905 at the five interior dates. Table \ref{tab:iv_mc_extended} compares the simultaneous-inference variants.

\begin{table}[htbp]
\centering
\caption{Extended Monte Carlo calibration of projected AR inference}
\label{tab:iv_mc_extended}
\footnotesize
\begin{tabularx}{\textwidth}{Xrrrr}
\hline\hline
Procedure & Pointwise coverage & Simultaneous coverage & Failure rate & Median critical \\
\hline
No CCE; parent resampling & 0.913 & 0.773 & 0.000 & 9.803 \\
LOO CCE; parent resampling & 0.911 & 0.778 & 0.000 & 9.806 \\
LOO CCE; parent/time-block resampling & 1.000 & 1.000 & 0.000 & 9.854 \\
LOO CCE; CD-selected parent/time resampling & 0.958 & 0.980 & 0.000 & 17.263 \\
LOO CCE; parent bootstrap/Bonferroni safeguard & 0.911 & 0.901 & 0.000 & 12.452 \\
\hline
\end{tabularx}
\begin{minipage}{0.96\textwidth}
\footnotesize
\textit{Notes:} Results use 1,000 replications and nominal 90\% inference. Pointwise coverage is averaged over seven evaluation dates; simultaneous coverage requires the entire seven-date family to cover. LOO denotes leave-one-parent-out. CD is the AR-moment cross-sectional-dependence diagnostic. Every procedure has a zero numerical-failure rate.
\end{minipage}
\end{table}

Pointwise calibration alone is not sufficient for a time path. Parent resampling with or without CCE gives pointwise coverage near 0.90 but simultaneous coverage of only 0.773 and 0.778. Applying parent/time-block resampling in every replication raises both coverage measures to 1.000 and is therefore markedly overconservative. The diagnostic-selected procedure is also conservative, with pointwise coverage 0.958, simultaneous coverage 0.980, and a median critical value of 17.263. The leading Bonferroni-safeguarded parent bootstrap produces pointwise coverage 0.911 and simultaneous coverage 0.901, with median critical value 12.452 and no numerical failures. It therefore provides the closest calibration to the nominal familywise level among the reported procedures.

These simulations establish a limited methodological result. Under the stated simulated data-generating processes, parent aggregation and projected AR inversion can control coverage, and the Bonferroni safeguard materially improves simultaneous calibration. The simulations do not add information to the observed first stage, test the empirical exclusion restriction, or transform all-real empirical confidence sets into informative ones. The empirical instrument is much closer to the weak core design than to the strong extended design. Monte Carlo evidence therefore validates aspects of the procedure, not the proposed instrument or a causal interpretation of the main estimates.

Taken together, the empirical diagnostics and simulation evidence clarify how the IV exercise bears on the main estimates. The exercise prevents a causal interpretation from being forced onto weak identifying variation. Weakness is present before localization, persists across bandwidths, and is not repaired by the alternative designs, crisis designs, generated first stages, penalization, or subgroup searches. Weak-IV-robust inference remains uninformative about causal direction and magnitude. Accordingly, the baseline, conditional, joint, sectoral, and aggregate time-varying estimates in the main text are interpreted as conditional associations. A causal investment effect would require an accessible instrument with materially stronger relevance and exclusion restriction.

\newpage

\section{Data}    \label{app:data}

\begin{table}[htbp]
\caption{Compustat and ExecuComp fields used to construct the firm-level empirical variables}
    \begin{tabularx}{\textwidth}{p{5cm}X}
        \hline\hline
        Vendor field        & Description                                 \\
        \hline
        \multicolumn{2}{l}{Compustat}                                     \\
        \hline
        gvkey               & Global Company Key                          \\
        at                  & Assets - Total                              \\
        dlc                 & Debt in Current Liabilities - Total         \\
        dltt                & Long-Term Debt - Total                      \\
        capx                & Capital Expenditures                        \\
        ppent               & Property, Plant and Equipment - Total (Net) \\
        che                 & Cash and Short-Term Investments             \\
        sale                & Sales/Turnover (Net)                        \\
        csho                & Common Shares Outstanding                   \\
        prcc\_f             & Price Close - Annual - Fiscal               \\
        seq                 & Stockholders Equity - Parent                \\
        txditc              & Deferred Taxes and Investment Tax Credit    \\
        \hline
        \multicolumn{2}{l}{ExecuComp}                                     \\
        \hline
        gvkey               & Company ID Number                           \\
        age                 & Executive's Age                             \\
        ceoann              & Annual CEO Flag                             \\ 
        becameceo           & Date Became CEO                             \\
        \hline
    \end{tabularx}
    \label{tab:compustat_and_execucomp_variables}
\end{table}

\begin{figure}
    \centering
    \includegraphics[width=0.5\linewidth]{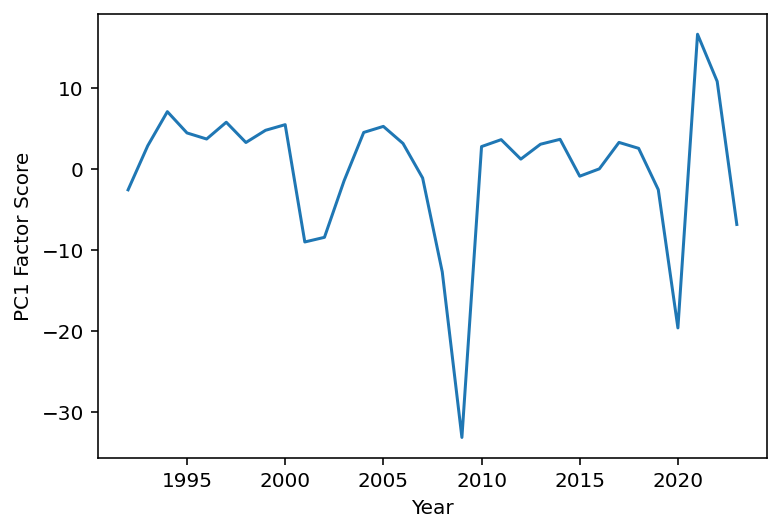}
    \caption{First principal component estimated from the FRED-QD macro dataset}
    \label{fig:pc1}
\end{figure}

\begin{figure}[p]
    \centering
    \includegraphics[width=0.95\textwidth]{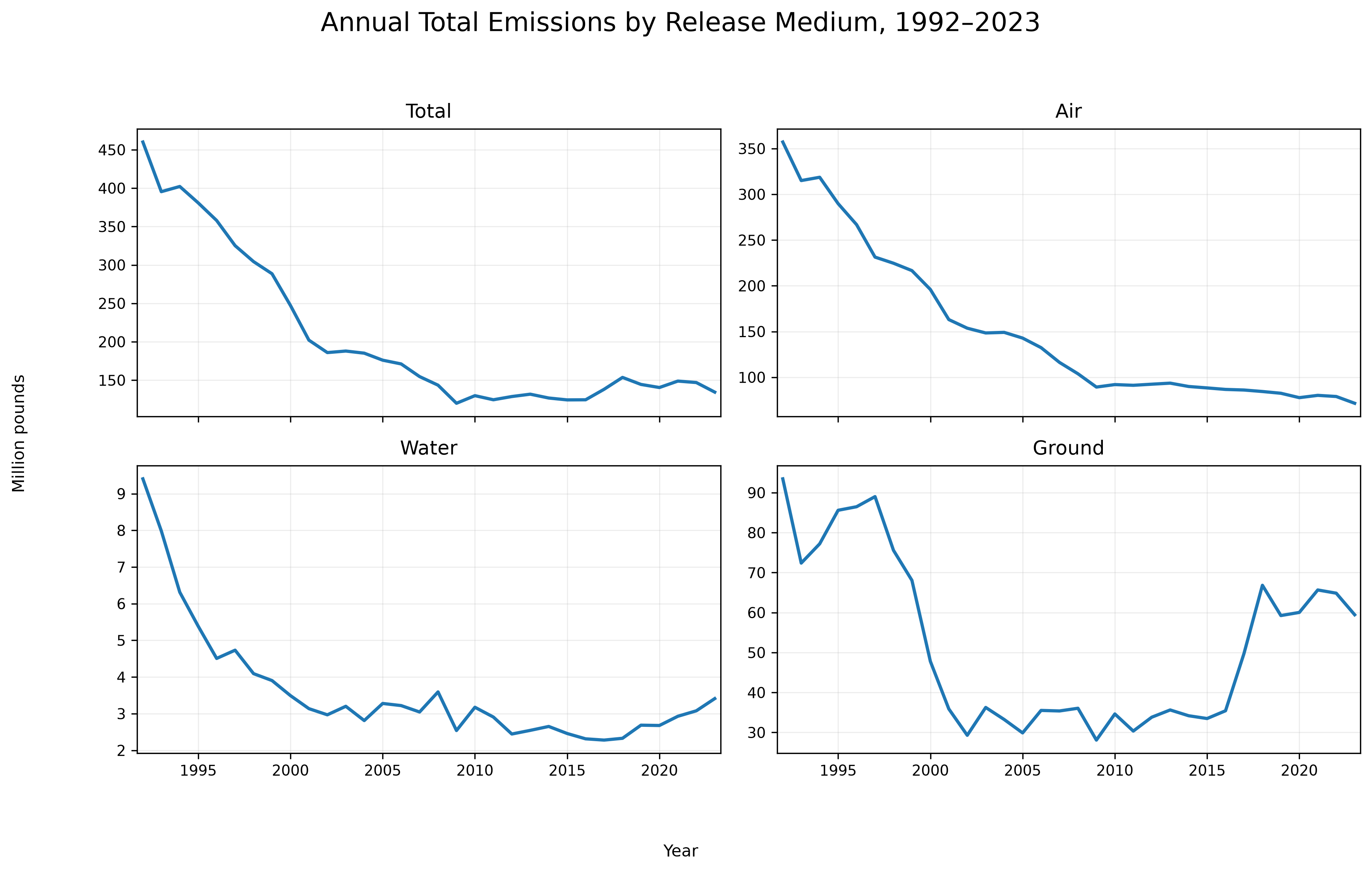}
    \caption{Annual aggregate facility emissions by release medium.}
    \label{fig:agg_emissions}
\end{figure}

\FloatBarrier

Transformations are needed to make variable series in the FRED-QD dataset stationary. ``Tcode" indicates different transformations required. With $X_t$ representing the original series, transformed type $\tilde{X}_t$ follows:
\begin{equation}
    \tilde{X}_t = 
    \begin{cases}
        X_t & \quad \text{if} \;\; \text{Tcode = 1} \\
        \Delta X_t & \quad \text{if} \;\; \text{Tcode = 2} \\
        \Delta^2 X_t & \quad \text{if} \;\; \text{Tcode = 3} \\
        log(X_t) & \quad \text{if} \;\; \text{Tcode = 4} \\
        \Delta log(X_t) & \quad \text{if} \;\; \text{Tcode = 5} \\
        \Delta^2 log(X_t) & \quad \text{if} \;\; \text{Tcode = 6} \\
        \Delta(X_t/X_{t-1} - 1) & \quad \text{if} \;\; \text{Tcode = 7} \\
    \end{cases}
    \label{equ:tcode}
\end{equation}

Detailed macroeconomic variables and corresponding T-codes are available at \citet{mccracken2020fred}.

\end{appendices}

\end{document}